\documentclass[fleqn,usenatbib]{mnras}

\usepackage{newtxtext,newtxmath}

\usepackage[T1]{fontenc}

\DeclareRobustCommand{\VAN}[3]{#2}
\let\VANthebibliography\thebibliography
\def\thebibliography{\DeclareRobustCommand{\VAN}[3]{##3}\VANthebibliography}

\usepackage{graphicx}	
\usepackage{amsmath}	
\usepackage{pdflscape}

\newcommand{\HI}{{\rm H}{\textsc i}}

\title[FEASTS: Gas and SF in interacting galaxies]{FEASTS: The fate of gas and star formation in interacting galaxies}

\author[S. Wang et al.]{
Shun Wang,$^{1}$
Jing Wang,$^{1}$\thanks{E-mail: jwang\_astro@pku.edu.cn}
Karen Lee-Waddell,$^{2,3,4}$
Dong Yang,$^{1}$
Xuchen Lin,$^{1}$
and Lister Staveley-Smith$^{2,5}$
\\
$^{1}$Kavli Institute for Astronomy and Astrophysics, Peking University, Beijing 100871, China\\
$^{2}$International Centre for Radio Astronomy Research (ICRAR), The University of Western Australia, 35 Stirling Highway, Crawley, WA 6009, Australia\\
$^{3}$CSIRO Space \& Astronomy, PO Box 1130, Bentley, WA 6102, Australia\\
$^{4}$International Centre for Radio Astronomy Research (ICRAR), Curtin University, Bentley, WA 6102, Australia\\
$^{5}$ARC Centre of Excellence for All-Sky Astrophysics in 3 Dimensions (ASTRO 3D), Australia
}

\date{Accepted XXX. Received YYY; in original form ZZZ}

\pubyear{\the\year{}}

\begin{document}
\label{firstpage}
\pagerange{\pageref{firstpage}--\pageref{lastpage}}
\maketitle

\begin{abstract}
We use \HI{} data from the FAST Extended Atlas of Selected Targets Survey (FEASTS) to study the interplay between gas and star formation of galaxies in interacting systems.
We build control and mock \HI{} disks and parameterize \HI{} disorder by a series of disorder parameters, describing the piling, clumpiness and expansion of \HI{}.
We find that interacting galaxies have higher \HI{} disorder described by almost all disorder parameters. Systems with comparable stellar masses and small relative velocities tend to have stronger expansion and clumpiness of \HI{}. At a given stellar mass, decreased \HI{} and total neutral gas mass and suppressed star formation rate of secondary galaxies are correlated with most disorder parameters. For primary galaxies, \HI{} and total neutral gas deficiency correlate with more \HI{} piling at two ends of the system outside \HI{} disks but not with the expansion or clumpiness of \HI{}. We also find that the \HI{} surface densities of both primary and secondary galaxies are lower within the \HI{} disks and higher outside compared to the control galaxies.
Our results suggest that while all the disorder parameters quantify the interaction strength almost equally well, they have different sensitivities in tracing star formation rate and gas mass enhancements. They also imply that while gas removal likely dominates the tidal effects on secondary galaxies, primary galaxies experience more complex situation that are possibly related to gas depletion and accretion happening at different interaction stages.
\end{abstract}

\begin{keywords}
galaxies: interactions -- galaxies: ISM -- galaxies: intergalactic medium -- galaxies: star formation -- galaxies: evolution
\end{keywords}



\section{introduction} \label{sec:intro}
Galaxies in our Universe are not ideally isolated. A large fraction of them reside in groups and clusters with neighbors at low redshift \citep[e.g.,][]{2013MNRAS.432..336W}.
In groups and clusters, central galaxies accrete less massive infalling galaxies. And satellites galaxies themselves also interact and even merge with each other.
The $\Lambda$-Cold Dark Matter ($\Lambda$CDM) cosmology predicts that the halo-halo merger is one of the major channel of mass and structure assembly.
The continuous merger of massive halos throughout their assembly history further increases the frequency of galaxy-galaxy interaction inside the halos \citep[e.g.,][]{1998ApJ...509..587F, 2003ApJ...582..141G}.
Tidal interaction between galaxies are thus inevitable and important for understanding the evolution of galaxies in the local Universe.

The effects of tidal interaction have been extensively studied in the literature.
Observational studies have focused on: 1) enhanced star formation and its dependence on galaxy separation \citep[e.g.,][]{2000ApJ...530..660B, 2008AJ....135.1877E}, mass ratio \citep[e.g.,][]{2007AJ....134..527W, 2012A&A...539A..45L}, primary/secondary identity \citep[e.g.,][]{2003MNRAS.346.1189L, 2007AJ....134..527W}, morphology \citep[e.g.,][]{2016ApJS..222...16C, 2018ApJS..237....2Z}, star formation activity \citep[e.g.,][]{2007AJ....134..527W, 2019ApJ...882...14M}, gas fraction \citep[e.g.,][]{2015MNRAS.449.3719S}, environment \citep[e.g.][]{2004MNRAS.352.1081A, 2010MNRAS.407.1514E}, and multiplicity \citep[e.g.,][]{2020MNRAS.494.3469B, 2021MNRAS.501.2969G}, 2) induced gas inflow deduced from metallicity measurements \citep[e.g.,][]{2012MNRAS.426..549S, 2013MNRAS.435.3627E}, 3) perturbed atomic hydrogen (\HI{}) \citep[e.g.,][]{2016MNRAS.459.1827P, 2019MNRAS.484..582B} and whether it is suppressed or enhanced \citep[e.g.,][]{2015MNRAS.448..221E, 2018ApJS..237....2Z, 2018MNRAS.478.3447E, 2022ApJ...927...66W}, 4) molecular hydrogen (H$_2$) content \citep[e.g.,][]{2018ApJ...868..132P, 2018MNRAS.476.2591V} and, 5) the effect of AGN activity \citep[e.g.,][]{2011MNRAS.418.2043E} and stellar feedback \citep[e.g.,][]{2019AJ....158..169S}.
More recent observations with improved spatial resolution present a more detailed picture \citep[e.g.,][]{2019MNRAS.482L..55T}. The preference of central star formation enhancement is confirmed, though with a dependence on the stage of interaction \citep[e.g.,][]{2019ApJ...881..119P}. The H$_2$ distribution also tends to be more centrally peaked when the galaxy is experiencing tidal interaction \citep[e.g.,][]{2019MNRAS.484.5192C}.

In particular, there are ongoing debates on how tidal interaction affects the gas content in different populations.
In statistical samples of interacting galaxies, e.g., those selected from SDSS, there is no universal agreement regarding the tidal effect on their \HI{} content.
\citet{2022ApJ...934..114Y} reported a marginal decrease in \HI{} gas fraction in major-merger pairs. \citet{2015MNRAS.448..221E} found no evidence of \HI{} deficiency in post-mergers with ALFALFA \citep{2018ApJ...861...49H} data. \citet{2018MNRAS.478.3447E} found that the \HI{} detection rate of post-merger galaxies is significantly higher than that of xGASS  \citep{2018MNRAS.476..875C}. And the median \HI{} gas fraction is $\sim 0.5$ dex higher than control galaxies with matched stellar masses. \citet{2018ApJS..237....2Z} also found that the \HI{} gas fractions of spiral galaxies in close galaxy pairs at $z = 0$ are consistent with the average of spiral galaxies in general.
Similar enrichment is also observed for molecular gas as a $0.4$ dex enhancement in gas fraction \citep{2018MNRAS.476.2591V} and a significant enhancement in gas mass \citep{2018ApJ...868..132P}.

In the specific interacting systems of Hickson compact groups (HCGs, \citealt{1982ApJ...255..382H}), on the other hand, \HI{} deficiency was ubiquitously found \citep[e.g.,][]{1987ApJS...63..265W, 1997A&A...325..473H}.
Such deficiency was confirmed by \citet{2012A&A...540A..96M} with a redshift-limit sample of HCG galaxies. \citet{2017MNRAS.464..957H} further found that the \HI{} deficiency of HCG 44 is not accounted for by additional \HI{} gas in the IGM environment revealed by deeper observation, tentatively confirming the speculation of \citet{2001A&A...377..812V} that phase transition in the IGM environment following the tidal stripping is responsible for the \HI{} deficiency. On the other hand, \citet{2019A&A...632A..78J} found that HCG 16 as a whole group is not \HI{} deficient after including \HI{} in the IGM environment, suggesting that the evolutionary track of \HI{} in interacting groups may not be unique \citep{2001A&A...377..812V}.
The molecular gas content in HCG galaxies shows more diversity. It can either be increased \citep[e.g.,][]{2012A&A...540A..96M}, decreased \citep[e.g.,][]{2001A&A...377..812V} or remain normal \citep[e.g.,][]{1998ApJ...497...89V}.

The diversity of gas deficiency of interacting galaxies may originate from the different strengths or evolutionary stage of the tidal interaction, different initial conditions for the galaxies and pairs/groups, etc.
The limitations of different types of observational data may also play a part.
Recently, \citet{2023ApJ...944..102W} separate a distinct phase of \HI{}, i.e., diffuse \HI{}, in the interacting NGC4631-NGC4656 system. By comparing FAST data to interferometry data, the authors identified a significant amount of diffuse \HI{} that was previously missed. The flux is missing partly due to the observational sensitivity, but mostly due to the missing zero-spacing in interferometers.
Further analysis reveals a close connection between the diffuse \HI{} and IGM. It tends to survive photon ionization and thermal evaporation and induce cooling out of hot IGM. So, the replenishment of ISM by cooling/accretion from IGM via diffuse \HI{} may be directly observed by FAST in such interacting systems. This may be a key to understanding \HI{} enrichment and deficiency in interacting systems.

Thus, we seek to understand the tidal effects on \HI{} content and star formation by studying the interacting galaxy pairs observed in FEASTS (FAST Extended-Atlas-of-Selected-Targets Survey) sample \citep{2023ApJ...944..102W}. These are simpler cases of tidal interaction than those in compact groups. Benefiting from the powerful FAST instrument, we minimize the possible observational effects introduced by limited sensitivity and the problem of missing flux that affects interferometric data. We are also able to well resolve the \HI{} component in these interacting systems so that the spatial distribution of \HI{} can be analyzed.

Our galaxy samples and data are introduced in Section~\ref{sec:sample}.
In Section~\ref{sec:ana} we demonstrate our method of obtaining all the measurements of gas and star formation.
The main results of our study are presented in Section~\ref{sec:result}. We discuss the implications for the fate of gas and star formation in interacting systems in Section~\ref{sec:dis}. Finally, in Section~\ref{sec:con} we summarize our main results.

\section{sample and data} \label{sec:sample}
\subsection{Sample galaxies from FEASTS} \label{sec:samp}
FEASTS is a $21$-cm emission line survey to map the extended \HI{} in Local Volume galaxies \citep{2023ApJ...944..102W, 2024ApJ...968...48W}. It takes advantage of the high sensitivity and low sidelobe levels of FAST (Five-hundred-meter Aperture Spherical radio Telescope) in China. It maps the extended and low-surface density \HI{} that is typically missed in interferometric observations, studying the evolution of Local Volume galaxies that have grand \HI{} disks, examining the properties of their ISM, IGM and the related physical connections and processes.

\begin{figure*}
    \centering
    \includegraphics[width=0.49\textwidth]{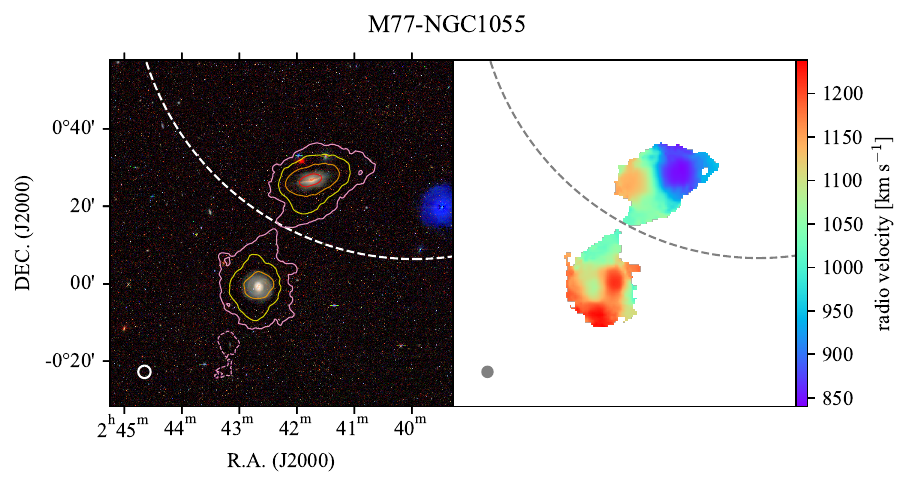}
    \includegraphics[width=0.49\textwidth]{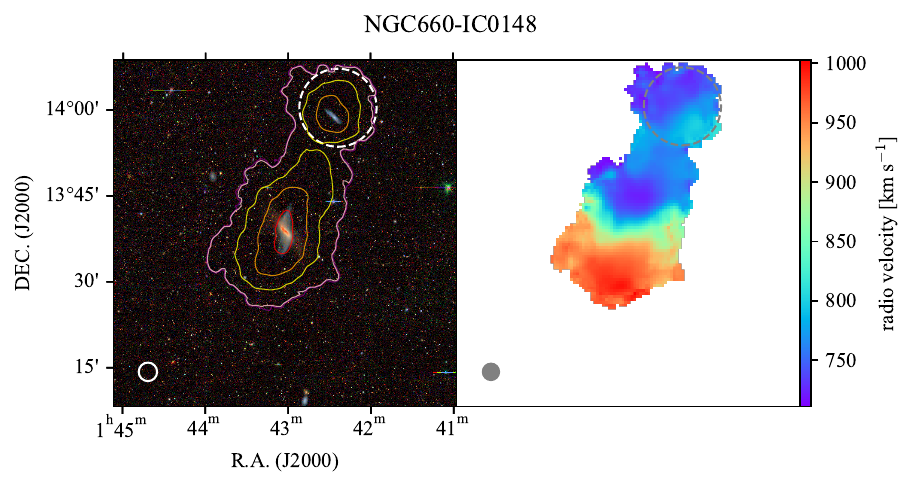}
    \includegraphics[width=0.49\textwidth]{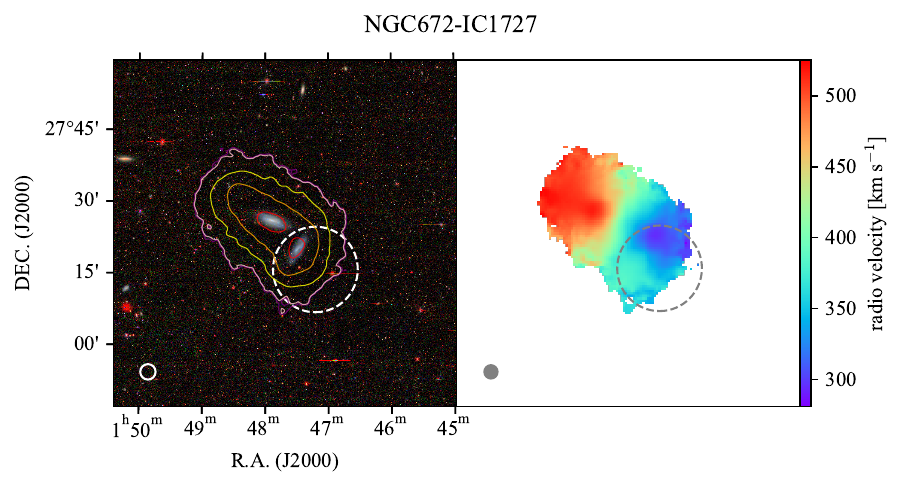}
    \includegraphics[width=0.49\textwidth]{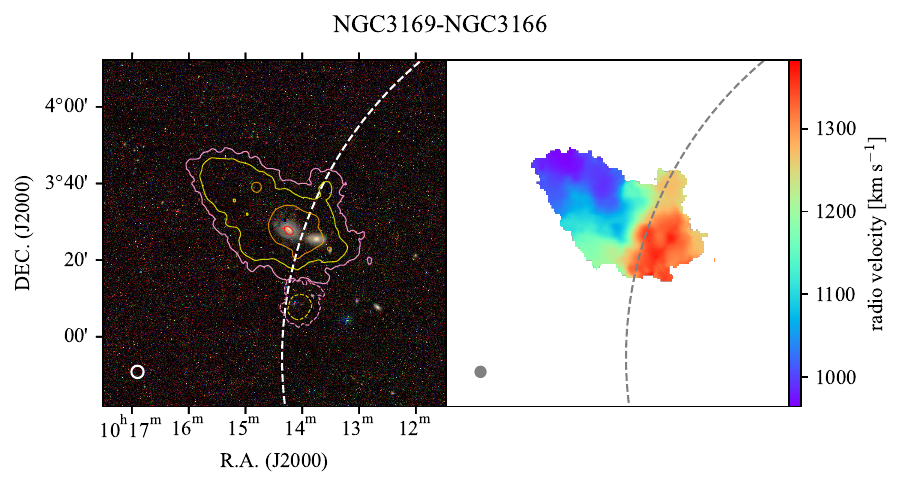}
    \includegraphics[width=0.49\textwidth]{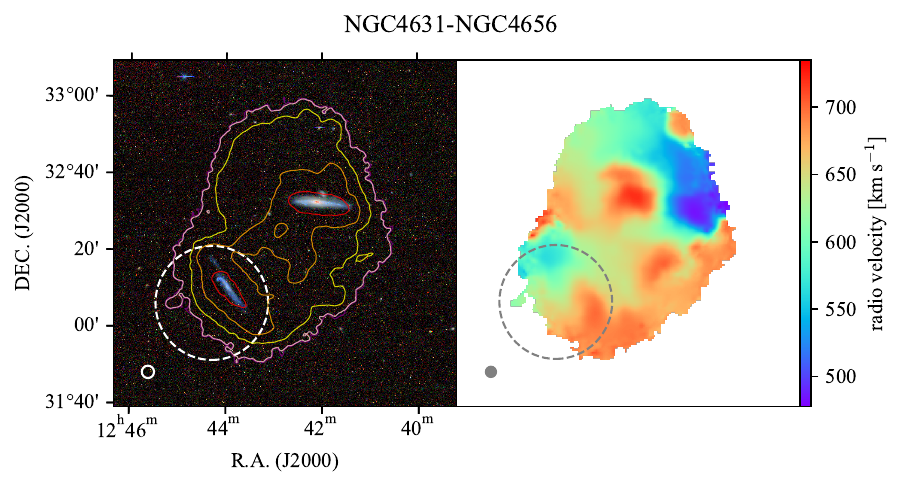}
    \includegraphics[width=0.49\textwidth]{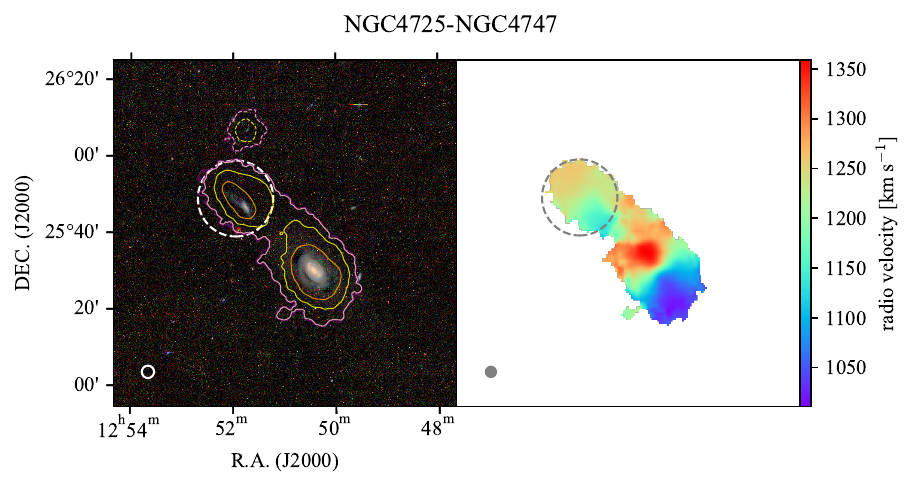}
    \includegraphics[width=0.49\textwidth]{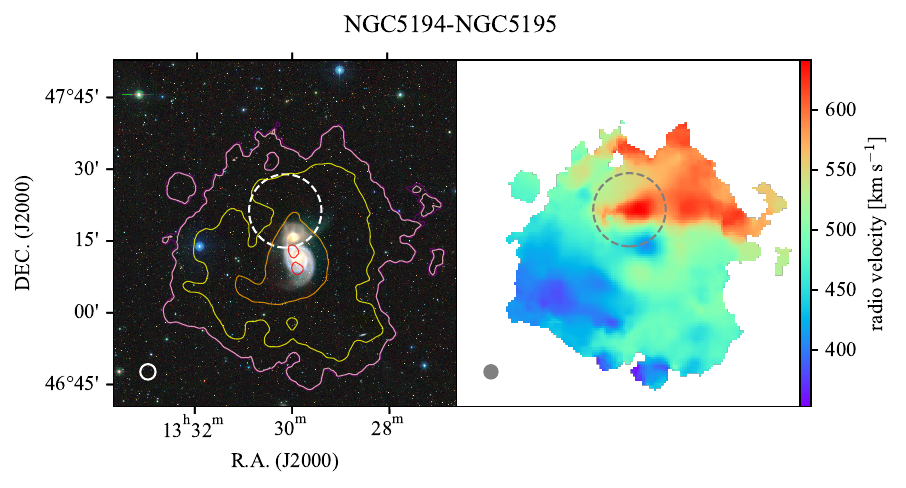}
    \includegraphics[width=0.49\textwidth]{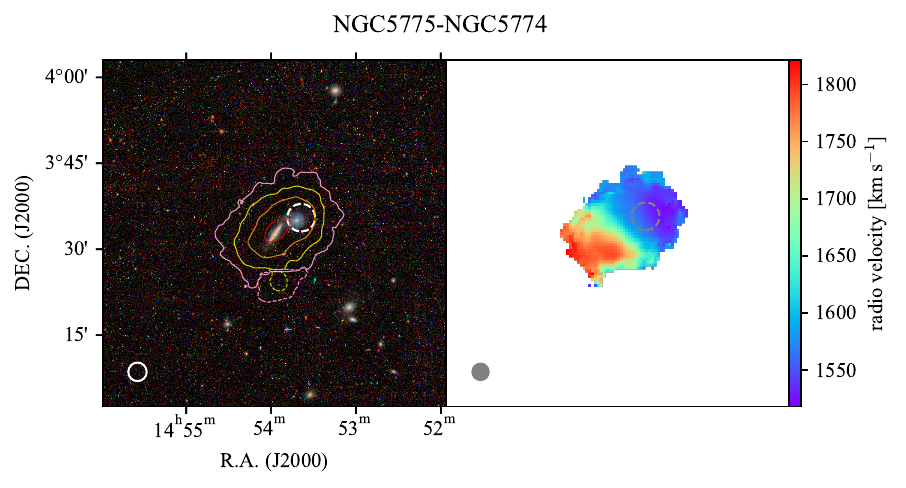}
\caption{Optical images with \HI{} column density contours overlaid and \HI{} moment-1 maps of the eight interacting systems. The optical images are from DESI Legacy Survey. The contour levels are $5-\sigma$ detection limit (purple, assuming $20$ km s$^{-1}$ line widths), $10^{18}$ (pink), $10^{19}$ (yellow), $10^{20}$ (orange) and $10^{21}$ (red) cm$^{-2}$. The solid contours enclose fluxes of the main \HI{} component of the systems. The dashed contours enclose \HI{} fluxes of small galaxies separated from the main \HI{} component of the systems (Section~\ref{sec:sourceid}). The white (in optical images) and gray (in moment-1 maps) dashed circles are the boundaries determined for \HI{} flux division between the primary and secondary galaxies (Section~\ref{sec:HIdiv}). \label{fig:mom01-int}}
\end{figure*}

We select eight interacting systems with prominent \HI{} tidal features and ten arbitrary isolated galaxies without optically-confirmed companions within a distance of $\sim 100$ kpc from the FEASTS sample. They occupy similar parameter space on \HI{} fraction main sequence \citep[HIMS,][]{2020MNRAS.493.1982J} and star forming main sequence \citep[SFMS,][]{2016MNRAS.462.1749S} as most primary galaxies of the interacting sample (Figure \ref{fig:MS}). There is only one primary galaxy, M77, having significantly higher SFR than the control galaxies for its $M_*$. We confirm that after removing the M77 associated galaxy pair, the control sample involved results in Section~\ref{sec:distur} and \ref{sec:eff} remain qualitatively the same. At the least, these isolated galaxies are well representative of normal-star-forming galaxies. These galaxies are listed in Table~\ref{tab:prop-int} and \ref{tab:prop-iso}. We present the moment-0 maps (column density images) and moment-1 maps (velocity fields) of the interacting systems in Figure~\ref{fig:mom01-int}. For NGC3169-NGC3166 and NGC4725-NGC4747 systems, we show the comparison between FEASTS channel maps and moment-0 maps with those from ALFALFA and GMRT observations \citep{2012MNRAS.427.2314L, 2016MNRAS.460.2945L} in Appendix~\ref{sec:AAGMRT}. It is obvious that previous GMRT observation missed a significant amount of large-scale \HI{} fluxes and diffuse \HI{} structures, while ALFALFA observations have higher noise level and worse spatial resolution.

\begin{landscape}
\begin{table}
\caption{Physical properties of galaxies in the interacting systems. Column (1): Identifier of the galaxies. The names of primary galaxies are in bold and the corresponding secondary galaxies are right below them. Column (2)-(3): R.A. and DEC. of the galaxy. Column (4): Heliocentric velocity of the galaxy. Column (5): $R_{\rm 25}$ of the galaxy. Column (6): Axis ratio of the galaxy, all from S$^4$G. Column (7): Position angle of the galaxy disk, defined as the angle of the semi-major axis in the approaching part of the disk, counting counter-clockwise from north, all from S$^4$G. Column (8): Inclination angle of the galaxy derived from axis ratio or cited from the reference noted. Column (9): Distance to the galaxy. Column (10): Log stellar mass of the galaxy. Column (11): Log star formation rate of the galaxy. Column (12): Log \HI{} mass of the galaxy. Column (13): Log H$_2$ mass of the galaxy, left blank if not available. Column (14): Log stellar mass ratio between secondary and primary galaxy. Column (15): Absolute relative velocity of the galaxy pair. Column (16): Projected distance between primary and secondary galaxy. \label{tab:prop-int}}
\begin{tabular}{crrrrrrrrrrrrlll}
\hline
{name} & {R.A. (J2000)} & {DEC. (J2000)} & {$V_{\rm hel}$}& {R$_{\rm 25}$} & {b/a} & {P.A.} & {incl.} & {Dist.} & {$\log M_*/{\rm M_\odot}$} & {log SFR/M$_\odot$ yr$^{-1}$} &  {$\log M_{\rm HI}/{\rm M_\odot}$} & {$\log M_{\rm H2}/{\rm M_\odot}$}  & {$\log M_{\rm *, sec} / M_{\rm *, prim}$} & {|$\Delta V_{\rm los}$|}&{$D_{\rm proj}$} \\ 
{} & {(deg)} & {(deg)} & {(km s$^{-1}$)}& {(arcsec)} & {} & {(deg)} & {(deg)} & {(Mpc)} & {} & {} & {} & {}  & {}& {(km s$^{-1}$)}&{(kpc)}\\ 
{(1)} & {(2)} & {(3)} & {(4)} & {(5)} & {(6)} & {(7)} & (8) & {(9)} & {(10)} & {(11)} & {(12)} & {(13)}  & {(14)}& {(15)}&{(16)}\\
\hline
\textbf{M77}   & 40.66962  & -0.01329  & 1137
& 212.40  & 0.998  & -80.0  & 4     & 12.30  & 10.71  & 1.51  & 9.10 &  6.53$^{d}$  & -0.27
& 141
&109.84 
\\
NGC1055 & 40.43847  & 0.44318  & 996
& 227.55  & 0.571  & 104.2  & 86$^{a}$ & 12.30  & 10.44  & 0.07  & 9.69 &  9.53$^{e}$  & 
& 
&
\\
\textbf{NGC672} & 26.97699  & 27.43278  & 429
& 217.35  & 0.455  & 68.4  & 65    & 8.28  & 9.37  & -0.68  & 9.49 &    -& -0.57
& 84
&19.49 
\\
IC1727 & 26.87454  & 27.33335  & 345
& 207.55  & 0.366  & 151.8  & 70    & 8.28  & 8.80  & -0.91  & 9.28 &    -& 
& 
&
\\
\textbf{NGC4631} & 190.53337  & 32.54151  & 610
& 464.65  & 0.247  & 87.1  & 85$^{b}$ & 7.30  & 10.05  & 0.30  & 10.06 &  9.19$^{f}$  & -1.22
& 36
&68.44 
\\
NGC4656 & 190.99037  & 32.17037  & 646
& 454.05  & 0.155  & 214.7  & 88    & 7.96  & 8.83  & -0.24  & 9.80 &    -& 
& 
&
\\
\textbf{NGC4725} & 192.61072  & 25.50076  & 1209
& 321.45  & 0.705  & 30.5  & 48    & 12.40  & 10.76  & -0.09  & 9.71 &  8.74$^{f}$  & -1.34
& 19
&87.97 
\\
NGC4747 & 192.93997  & 25.77503  & 1190
& 104.00  & 0.374  & 37.2  & 70    & 14.30  & 9.42  & -0.73  & 9.35 &    -& 
& 
&
\\
\textbf{NGC3169} & 153.56266  & 3.46609  & 1232
& 130.95  & 0.754  & 238.4  & 44    & 23.70  & 10.98  & 0.29  & 10.26 &  9.79$^{e}$  & -0.05
& 49
&53.15 
\\
NGC3166 & 153.44053  & 3.42481  & 1183
& 143.60  & 0.568  & 81.9  & 62    & 23.70  & 10.93  & -0.19  & 9.72 &  9.32$^{e}$  & 
& 
&
\\
\textbf{NGC660} & 25.75979  & 13.64568  & 848
& 249.55  & 0.390  & 183.6  & 76    & 9.00  & 9.79  & 0.23  & 9.54 &  9.31$^{e}$  & -1.62
& 81
&56.80 
\\
IC0148 & 25.61234  & 13.97703  & 767
& 101.65  & 0.239  & 227.8  & 79    & 9.00  & 8.17  & -1.53  & 8.70 &    -& 
& 
&
\\
\textbf{NGC5775} & 223.49022  & 3.54447  & 1676
& 125.05  & 0.511  & 145.0  & 86$^{c}$ & 19.80  & 10.47  & 0.46  & 9.96 &    -& -1.06
& 108
&25.41 
\\
NGC5774 & 223.42693  & 3.58249  & 1568
& 90.60  & 0.842  & 153.6  & 33    & 19.80  & 9.41  & -0.55  & 9.46 &    -& 
& 
&
\\
\textbf{NGC5194} & 202.46957  & 47.19526  & 460
& 336.60  & 0.656  & 175.7  & 51    & 8.58  & 10.73  & 0.65  & 9.66 &  9.46$^{g}$  & -0.37
& 5
&11.02 
\\
NGC5195 & 202.49829  & 47.26613  & 455& 172.65  & 0.708  & -62.7  & 49    & 8.58  & 10.36  & -0.16  & 8.99 &    -& & &\\
\hline
\end{tabular}\\
$^a$\citet{2014ApJ...795..136S}
$^b$\citet{2011AA...526A.118H}
$^c$\citet{1994ApJ...429..618I}
$^d$\citet{2003ApJS..145..259H}
$^e$\citet{2019PASJ...71S..14S}
$^f$\citet{2012AJ....143..138S}
$^g$\citet{2008AJ....136.2782L}
\end{table}
\end{landscape}

\begin{table*}
\caption{Physical properties of the isolated galaxies. Column (1): Identifier of the galaxies. Column (2)-(3): R.A. and DEC. of the galaxy. Column (4): Axis ratio of the galaxy from S$^4$G, or Simbad where noted. Column (5): Position angle of the galaxy disk, defined as the angle of the semi-major axis of the approaching part of the disk, counting counter-clockwise from north. From S$^4$G, or Simbad where noted. Column (6): Inclination angle of the galaxy derived from axis ratio, or cited from the reference noted. Column (7): Distance to the galaxy. Column (8): Log stellar mass of the galaxy. Column (9): Log \HI{} mass of the galaxy. Column (10): Log star formation rate of the galaxy. \label{tab:prop-iso}}
\begin{tabular}{crrrrrrrrr}
\hline
{name} & {R.A. (J2000)} & {DEC. (J2000)} & {b/a} & {P.A.} & {incl.} & {Dist.} & {$\log M_* / {\rm M_\odot}$} & {$\log M_{\rm HI} / {\rm M_\odot}$}  &{log SFR/M$_\odot$ yr$^{-1}$}\\ 
{} & {(deg)} & {(deg)} & {} & {(deg)} & {(deg)} & {(Mpc)} & {} & {}  &{}\\ 
{(1)} & {(2)} & {(3)} & {(4)}& {(5)}& {(6)}& {(7)}& {(8)}& {(9)} &{(10)}\\
\hline
NGC891 & 35.63711  & 42.34832  & 0.199$^{c}$  & 202.0$^{c}$  & 84$^{a}$ & 9.91  & 10.72  & 9.72   &0.32
\\
NGC4517 & 188.18994  & 0.11504  & 0.178  & 261.6  & 84    & 11.10  & 10.22  & 9.66   &-0.15
\\
NGC3344 & 160.87979  & 24.92222  & 0.900  & -58.6  & 27    & 10.00  & 10.29  & 9.75   &-0.08
\\
NGC4559 & 188.99017  & 27.95996  & 0.421  & -37.0  & 67    & 8.91  & 9.81  & 9.83   &-0.18
\\
NGC7331 & 339.26688  & 34.41578  & 0.398$^{c}$  & 171.0$^{c}$  & 76    & 14.70  & 11.00  & 10.13   &0.53
\\
NGC2903 & 143.04213  & 21.50083  & 0.492  & 199.4  & 63    & 8.47  & 10.42  & 9.66   &0.32
\\
NGC628 & 24.17394  & 15.78364  & 0.840  & 118.0  & 34    & 9.77  & 10.24  & 10.11   &0.23
\\
NGC925 & 36.82047  & 33.57888  & 0.549$^{c}$  & 282.0$^{c}$  & 61    & 9.16  & 9.75  & 9.82   &-0.17
\\
NGC3521 & 166.45237  & -0.03590  & 0.566  & -18.5  & 73$^{b}$ & 11.20  & 10.83  & 10.03   &0.42
\\
NGC4244 & 184.37358  & 37.80711  & 0.188  & 225.1  & 90$^{a}$ & 4.29  & 9.20  & 9.32   &-0.95\\
\hline
\end{tabular}\\
$^{a}${\citet{2011AA...526A.118H}}
$^{b}${\citet{2008AJ....136.2563W}}
$^{c}${from Simbad}
\end{table*}

\subsection{FEASTS HI data}
The observation setting of FEASTS is described in \citet{2023ApJ...944..102W, 2024ApJ...968...48W}, and we briefly summarize it here. The FEASTS \HI{} data is observed using the on-the-fly multi-beam scan mode of FAST. Each target is scanned in both RA and Dec directions for $\sim 1^\circ$ with at least three passes each. The scans were conducted with the $L-$band $19$ beam receiver rotated, so that the effective angular separation between scan lines is $1'.15$. The data is recorded using the \textit{Spec(W+N)} backend with a channel width of $7.63$ kHz, or $1.61$ km s$^{-1}$ for \HI{} $21$-cm observations in the nearby Universe.

Information about new FEASTS observations are shown in Table~\ref{tab:samp}. Observations of NGC628, NGC925, NGC2903, NGC3521, NGC4631, NGC5194 and NGC7331 are presented in \citet{2023ApJ...944..102W, 2024ApJ...968...48W}, and those of NGC891, NGC1055, NGC4244, NGC4517 and NGC5775 are presented in Yang et al. (in prep.), which are all part of the FEASTS program.

\begin{table*}
\caption{Basic information about new FEASTS observations of the sample systems and galaxies. Column (1): Identifier of the systems or galaxies. Column (2)-(3): R.A. and DEC. of the field center. Column (4): Size of the Field-of-View. Column (5): The date of observation. Column (6): The effective integration time per l-o-s. Column (7): The noise level of the data cube. Column (8): The corresponding detection limit assuming $20$ km s$^{-1}$ line widths and $3-\sigma$ signal. \label{tab:samp}}
\begin{tabular}{crrccccc}
\hline
{name} & {R.A. (J2000)} & {DEC. (J2000)} & {FoV} & {obs. date} & {int. time per beam} & {noise} & {depth} \\ 
{} & {(deg)} & {(deg)} & {(deg $\times$ deg)} & {(YYYYMMDD)} & {(s)} & {(mJy beam$^{-1}$)} & {($10^{-17}$ cm$^{-2}$)} \\
{(1)} & {(2)} & {(3)} & {(4)} & {(5)} & {(6)} & {(7)} & {(8)} \\
\hline
NGC672-IC1727 & 26.9211  & 27.3887  & 1.2$^\circ$ $\times$ 1.2$^\circ$ & 20221115 & 221.6  & 0.95  & 4.74  \\
NGC4725-NGC4747 & 192.7470  & 25.6636  & 1.5$^\circ$ $\times$ 1.5$^\circ$ & 20220329 & 229.0  & 0.89  & 4.50  \\
NGC3169-NGC3166 & 153.6215  & 3.4505  & 1.5$^\circ$ $\times$ 1.5$^\circ$ & 20220306 & 207.1  & 1.12  & 5.62  \\
NGC660-IC0148 & 25.7590  & 13.6420  & 1.0$^\circ$ $\times$ 1.0$^\circ$ & 20211115 & 214.4  & 0.95  & 4.77  \\
NGC3344 & 160.8796  & 24.9237  & 1.2$^\circ$ $\times$ 1.2$^\circ$ & 20220327 & 229.8  & 1.01  & 5.07  \\
NGC4559 & 188.9893  & 27.9614  & 1.2$^\circ$ $\times$ 1.2$^\circ$ & 20220425 & 235.9  & 1.01  & 5.07  \\
\hline
\end{tabular}
\end{table*}

The data reduction is carried out with a pipeline developed following the standard procedure of reducing radio single-dish data, particularly that from the Arecibo Legacy Fast ALFA Survey \citep{2018ApJ...861...49H} and \HI{} Parkes All Sky Survey \citep{2001MNRAS.322..486B}. More detailed description of the data reduction procedure is referred to \citet{2023ApJ...944..102W}.

\subsection{Archival H2 data} \label{sec:CO}
We use the CO observation from HERACLES \citep{2009AJ....137.4670L}, BIMA \citep{2003ApJS..145..259H} and COMING \citep{2019PASJ...71S..14S} to derive the molecular gas distribution of the galaxies in our interacting systems. Among $16$ galaxies in the interacting sample, we have three, one and four CO images of galaxies from HERACLES, BIMA and COMING, respectively (Table~\ref{tab:prop-int}).

\subsection{Ancillary information} \label{sec:ancc}
The positions of the galaxies ($R.A.$ and $DEC.$) are from the Simbad database and the heliocentric velocities are from the NED database.
The optical disk sizes R$_{\rm 25}$ are also from the NED database, which are all from the RC3 catalog \citep{1991rc3..book.....D}.
The galaxy $T-Type$\footnote{In cases where $T-type$ is from Simbad, we translate `SB', `SAB' and `SA' into $T-type = 3, 2, 1$, respectively.}, axis ratio and position angle of the optical disks are from the S$^4$G \citep{2010PASP..122.1397S} database when available, and Simbad database otherwise.
Inclinations of the galaxies are from various literature (see references in Table~\ref{tab:prop-int} and Table~\ref{tab:prop-iso}) when available. Otherwise they are derived from the axis ratio and morphological type following  \citet{1990ApJ...349....1F}. We remind the readers to be cautious about the inhomogeneity in these parameters.
Distance, stellar mass and star formation rate (SFR) are from the Z0MGS \citep{2019ApJS..244...24L} catalog.
These physical properties of the interacting and isolated galaxies are shown in Table~\ref{tab:prop-int} and Table~\ref{tab:prop-iso}, respectively.

\section{analysis} \label{sec:ana}
\subsection{Data preparation}
\subsubsection{Source identification in HI cubes} \label{sec:sourceid}
We use \textit{SoFiA} \citep{2015MNRAS.448.1922S} to detect \HI{} emissions from the data cube and generate source masks. The threshold-based smooth$+$clipping source-finding algorithm is used with the reliability module enabled. More technical details are referred to \citet{2023ApJ...944..102W}. The produced source masks are then used to generate moment-0, moment-1 maps and derive the noise level of each data cube. We manually check and confirm there are no floating \HI{} clumps or other tidal features that may originate from interaction outside the selected sources.

Following the procedure of \citet{2023ApJ...944..102W} and Huang et al. (in prep.), we apply the watershed algorithm to the moment-0 maps to separate the \HI{} fluxes of small galaxies from the main galaxies of the interacting pairs (Appendix~\ref{sec:watershed}). We confirm that including the flux in these pixels does not significantly change the results presented in this work.

\subsubsection{Division of HI flux between primary and secondary galaxies} \label{sec:HIdiv}
For each interacting system, we denote the galaxy that has larger stellar mass as primary galaxy and the other as secondary galaxy.
We divide the \HI{} fluxes of the two galaxies by deciding a 2D boundary upon which a test mass experiences equal strength of tidal torque influence \citep{1991A&A...244...52E} from them. The boundary is a circle of Apollonius on the sky plane defined by center $C$ (Equation~\ref{equ:tidal1}) and radius $R$ (Equation~\ref{equ:tidal2}). The \HI{} flux inside the circular boundary is assigned to the secondary galaxy and the flux outside is assigned to the primary.

\begin{equation} \label{equ:tidal1}
C = \left( \frac{X_{\rm prim} - k^2X_{\rm sec}}{1 - k^2}, \frac{Y_{\rm prim} - k^2 Y_{\rm sec}}{1 - k^2} \right)
\end{equation}
\begin{equation} \label{equ:tidal2}
R = \frac{k}{1 - k^2} d_{\rm prim2sec}
\end{equation}
\begin{equation} \label{equ:tidal3}
k \equiv \frac{d_{\rm prim}}{d_{\rm sec}} = \left( \frac{M_{\rm prim}}{M_{\rm sec}} \right)^{\frac{1}{3}}
\end{equation}
$(X_{\rm prim}, Y_{\rm prim})$ and $(X_{\rm sec}, Y_{\rm sec})$ are the coordinates of the two galaxies on the sky, $k$ is the ratio of the distances from any point on the circle to the two galaxies, and $d_{\rm prim2sec}$ is the distance between the two galaxies.

Note that there are multiple factors resulting in uncertainty in flux division, including projection effects and neglected orbital history of the system. And such division does not reflect the origin of gas, for \HI{} flux assigned to a galaxy can be either from its own \HI{} disk, the companion or even cooling induced in CGM. These uncertainties propagate to following analysis that is based on the \HI{} mass measurements performed here.

\subsection{Constructing comparison images}
\subsubsection{Construction of control HI disks} \label{sec:ctrl}
We use the \HI{} disks of isolated FEASTS galaxies (Section~\ref{sec:samp}) to construct control \HI{} disks as follows:

\begin{enumerate}
    \item  For each interacting galaxy, isolated galaxies with matched inclinations (either $incl. < 50^\circ$, $50^\circ < incl. < 80^\circ$ or $incl. > 80^\circ$) are selected.
    \item The \HI{} disks of the selected isolated galaxies are enlarged, so that the enlarged \HI{} disks have the same apparent size as that of the interacting galaxy.
    \item       The enlarged \HI{} disks are further rotated to match the position angles of the interacting galaxy. Two rotated disks can be produced based on one enlarged \HI{} disk (rotated by $180^\circ$).
    \item Finally, moment-0 maps of all possible combinations of the rotated \HI{} disks for interacting galaxies are produced.
\end{enumerate}

A number of $N_{\rm control} = 4 \times N_{\rm prim} \times N_{\rm sec}$ control moment-0 maps are produced for each system, where $N_{\rm id}$ ($id = prim, sec$) is the number of isolated galaxies in the corresponding inclination bin for the primary and secondary galaxy, respectively. $N_{\rm control}$ ranges from $24$ to $100$ for all the interacting systems in our sample.
We do not use \HI{} disks observed with interferometry in the literature, so that the possible effects of the missing flux problem are avoided.

\subsubsection{Construction of mock HI disks}
We use the \textit{GALMOD} module of $^{3D}$BAROLO \citep{2015MNRAS.451.3021D} to construct mock \HI{} disks to see how an `average' \HI{} disk would look like if observed under the same conditions. They will serve as reference or baseline images to be subtracted for highlighting and characterization of abnormal \HI{} morphology in the interacting system images and in the control images (Section~\ref{sec:disorder}). We note that, these averaged disks are based on previous characterization with the interferometry data, and thus do not necessarily represent the average of \HI{} disks detected in FEASTS, which will be studied with larger samples in the future (Wang et al. in prep).

The size of the \HI{} disk ($R_{\rm HI}$) is derived based on the \HI{} size-mass relation \citep{2016MNRAS.460.2143W}.
We linearly extrapolate the universal \HI{} radial profile out to $2 R_{\rm HI}$ following the extrapolation performed in \citet{2020ApJ...890...63W}, and use it as the radial distribution of the mock \HI{} disks (with a small adjustment, Appendix~\ref{sec:gauss}).
The scale height of the mock \HI{} disks is set as a function of distance to the disk center, going up linearly from $0.1$ kpc at the center to $0.6$ kpc at a radius of $20$ kpc and flattens at larger radii. Such functional form is determined as a simplified average of those derived in \citet{2019A&A...622A..64B, 2019A&A...632A.127B, 2020A&A...644A.125B}.

\begin{figure*}
    \centering
    \includegraphics[width=0.43\textwidth]{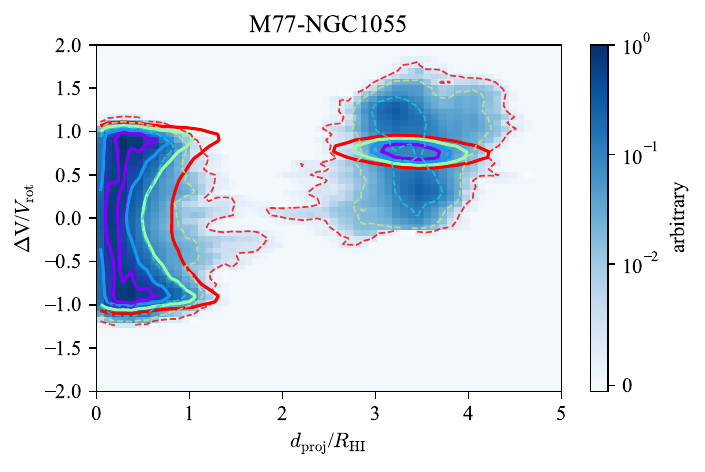}
    \includegraphics[width=0.43\textwidth]{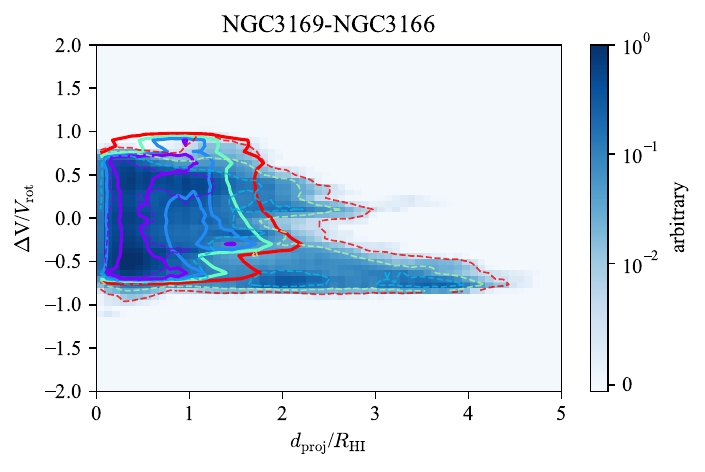}
    \includegraphics[width=0.43\textwidth]{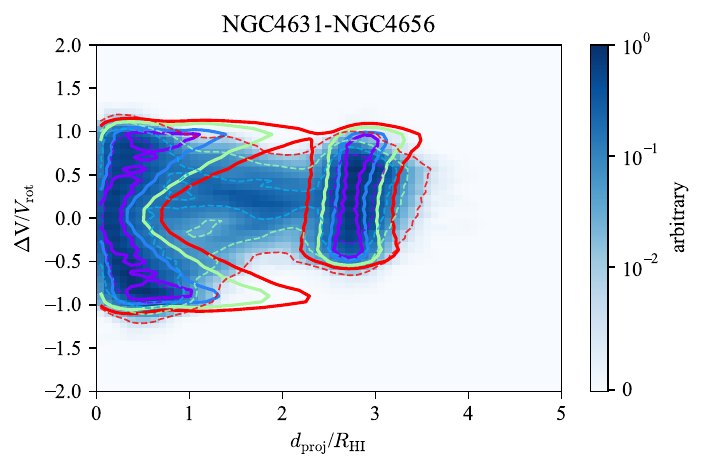}
    \includegraphics[width=0.43\textwidth]{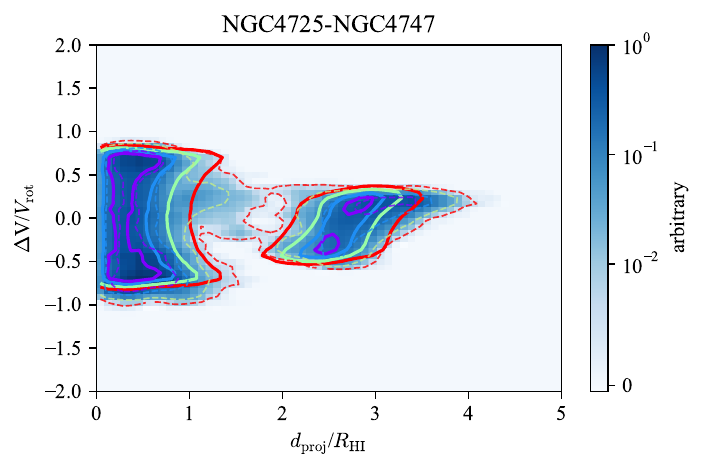}
    \includegraphics[width=0.43\textwidth]{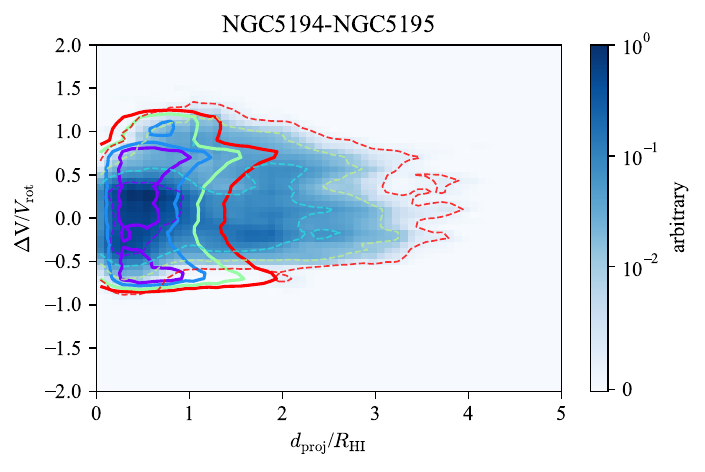}
    \includegraphics[width=0.43\textwidth]{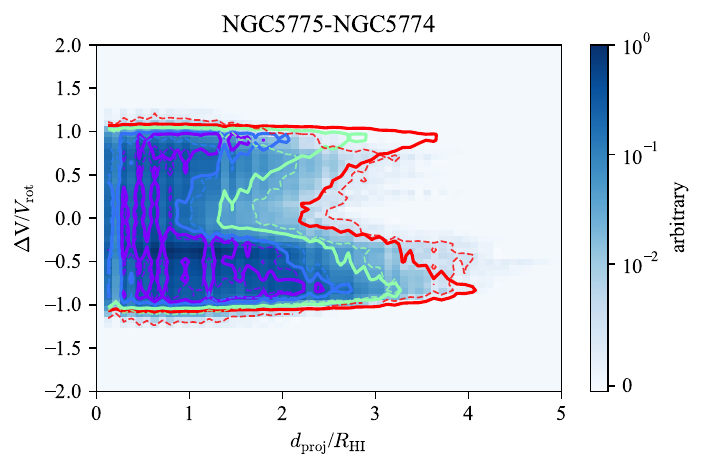}
    \includegraphics[width=0.43\textwidth]{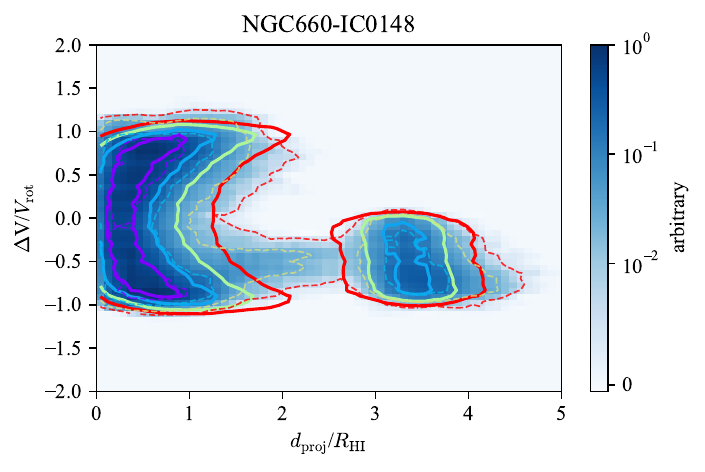}
    \includegraphics[width=0.43\textwidth]{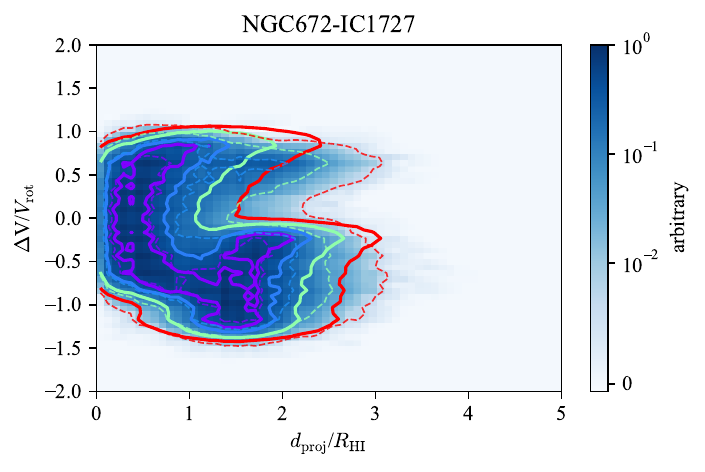}
\caption{\HI{} PSD of the interacting systems with contour of mock and real PSD overlaid. The projected distance $d_{\rm proj}$ is normalized by $R_{\rm HI}$ and the relative velocity $\Delta V$ is normalized by $V_{\rm rot}$ of the reference galaxy. The reference galaxy is the primary galaxy except for the NGC1055-M77 system where it is the secondary galaxy. The pixel sizes are determined individually for each system. The pixel values are all normalized so they peak at unity in each PSD. The solid and dashed contours enclose $99.7 \%$ (red), $97.5 \%$ (green), $84.1 \%$ (blue) and $50 \%$ (purple) fluxes of the mock and real PSD, respectively. \label{fig:PSD}}
\end{figure*}

The rotation curves of the mock \HI{} disks are assigned according to their baryonic masses. We divide SPARC galaxies \citep{2016AJ....152..157L} into five different baryonic mass bins and derive a median rotation curve out to the radius at which we still have measurements of the rotational velocities for more than half galaxies in each bin. The rotation curves are assumed to be flat beyond this range. The median rotation curves are smoothed before assigned to the mock \HI{} disks in the corresponding baryonic mass bin.
We additionally scale the rotation curve in some cases so that the \HI{} flux distribution of mock in PSD (Figure~\ref{fig:PSD}) matches the sample galaxies reasonably well.

$^{3D}$BAROLO then produces data cubes based on these input. After convolving the channel maps with the average FAST beam \citep{2023ApJ...944..102W}, we obtain the moment images and fluxes via the same procedure as the reduction of FEASTS data. Mock moment-0 maps are obtained for each interacting system by putting the two corresponding mock \HI{} disks on the same image. Similar mock maps are produced for the images of control galaxies. The mock \HI{} disks of control galaxies are enlarged and rotated in the same way as we treat the controls (Section~\ref{sec:ctrl}).
We note that mocks only represent the average, i.e., no scatter around this average is simulated (radial profile, scale height and rotation curve), which may give rise to systematic uncertainties in our sample. And the assumed \HI{} radial profile is derived from interferometry data, which may suffer from the missing flux problem \citep{2023ApJ...944..102W}.

We present \HI{} PSDs of the interacting galaxies in Figure~\ref{fig:PSD} with contours of the real and mock PSD overlaid. In all eight interacting systems, the \HI{} distribution show different level of deviation from that of the mock. In Appendix~\ref{sec:PSD} we present a detailed description of these \HI{} PSDs.

\subsection{Parametrization and measurements}
\subsubsection{Disorder parameters} \label{sec:disorder}
\begin{figure}
    \centering
    \includegraphics[width=0.75\linewidth]{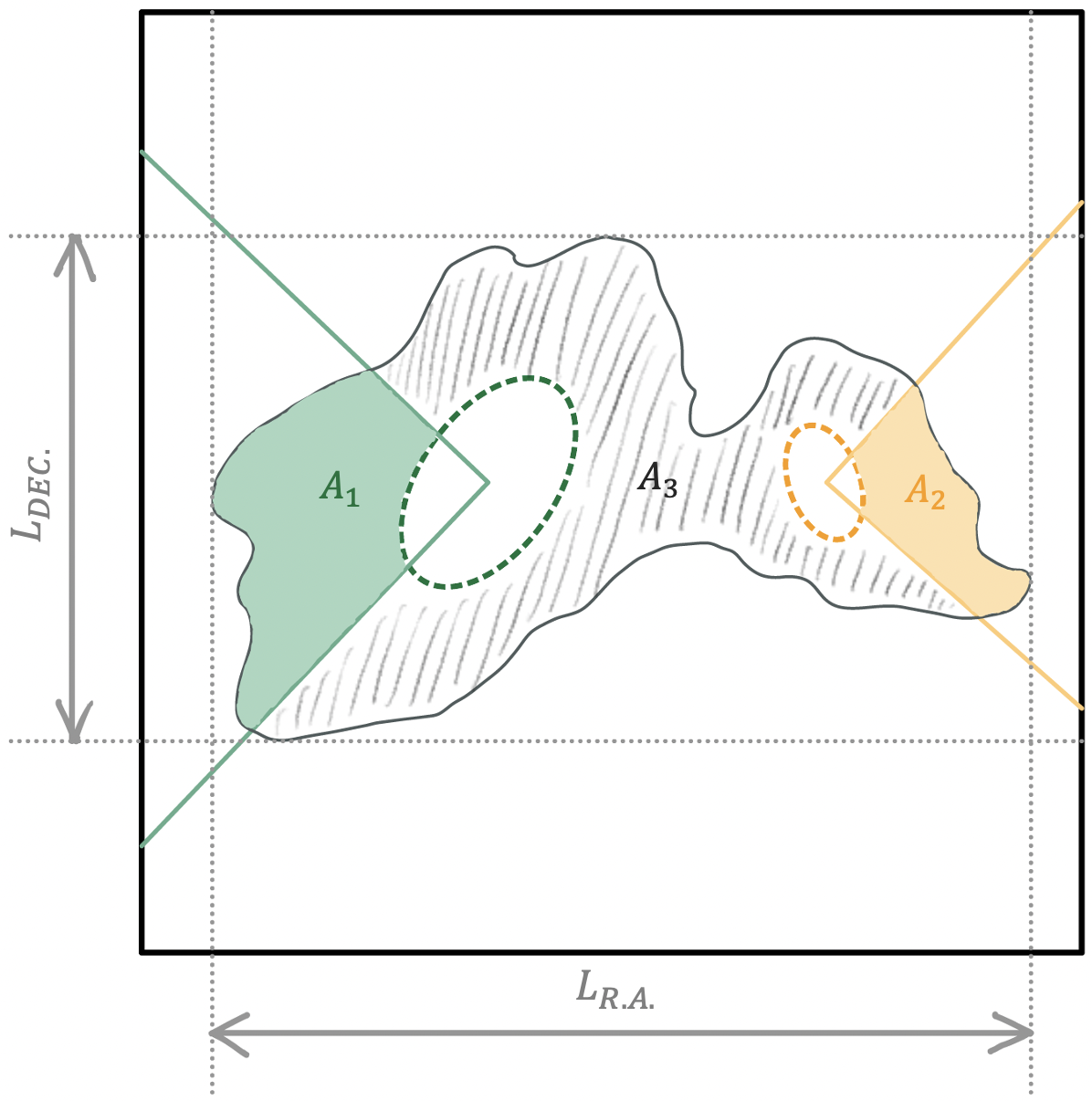}
\caption{A cartoon illustrating the definition of different regions. Symbols in green and orange are related to primary and secondary galaxy, respectively. \HI{} disks are shown as dashed ellipses. The solid lines enclose the regions at two ends of the system. Regions denoted as $A_1$ and $A_2$ are highlighted in green and orange, respectively. Region denoted as $A_3$ is hatched. The thin solid gray curve outlines the source mask produced by \textit{SoFiA}. \label{fig:A123}}
\end{figure}

We characterize disorder in the \HI{} morphologies using the moment-0 and the residual maps. The residual maps are constructed by subtracting mock moment-0 maps from the FEASTS and the control moment-0 maps. An illustration of the regions in the residual maps is shown in Figure~\ref{fig:A123}, which is produced as follows:

We firstly rotate the residual maps so that the primary galaxy is on the left of the secondary galaxy.
Next, we plot two $45^\circ$ lines extending from the center of the primary galaxy to the left, and two from the center of the secondary galaxy to the right.
Finally, we plot the contour of detected \HI{} fluxes at a column density of $10^{18}$ cm$^{-2}$. We also plot the \HI{} disk regions with semi-major axis set to $R_{\rm HI}$, and position angles and axis ratios determined by the optical photometry (Section~\ref{sec:ancc}).

The rotated residual maps of the interacting systems are presented in Section~\ref{sec:distur} and Figure~\ref{fig:res} therein. To quantify disorder in the \HI{} morphologies and enable further analysis of their dependence and impact on galactic properties, we measure the \HI{} flux $F$ in the moment-0 maps, the residual \HI{} flux $R$ and the fraction of positive pixels $f^+$ in the residual maps in the following regions:
\begin{itemize}
    \item The entire region labeled as detected sources in \textit{SoFiA} source mask (region enclosed by thin solid gray contour in Figure~\ref{fig:A123}). Quantities derived in this region are denoted by subscript $_{\rm tot}$.
    \item The region labeled as detected sources in \textit{SoFiA} source mask, with two \HI{} disks excluded ($A_1 \cup A_2 \cup A_3$ in Figure~\ref{fig:A123}). Quantities derived in this region are denoted by subscript $_{\rm 123}$.
    \item The region labeled as detected sources in \textit{SoFiA} source mask at two ends of the system, with two \HI{} disks excluded ($A_1 \cup A_2$ in Figure~\ref{fig:A123}). Quantities derived in this region are denoted by subscript $_{\rm 12}$.
\end{itemize}

Finally the following parameters are calculated, among which $\frac{R_{\rm 123}}{F_{\rm 123}}$, $\frac{F_{\rm 123}}{F_{\rm tot}}$ and $f^+_{\rm 123}$ quantify the level of \HI{} piling outside \HI{} disks, and $\frac{R_{\rm 12}}{F_{\rm 12}}$, $\frac{F_{\rm 12}}{F_{\rm tot}}$, $\frac{F_{\rm 12}}{F_{\rm 123}}$, $\frac{R_{\rm 12}}{F_{\rm 123}}$ and $f^+_{\rm 12}$ quantify the level of \HI{} piling at two ends of the system outside \HI{} disks:
\begin{itemize}
    \item $\frac{R_{\rm 123}}{F_{\rm 123}}$, the fraction of residual flux outside \HI{} disks.
    \item $\frac{R_{\rm 12}}{F_{\rm 12}}$, the fraction of residual flux at two ends of the systems.
    \item $\frac{F_{\rm 123}}{F_{\rm tot}}$, the fraction of fluxes outside the \HI{} disks.
    \item $\frac{F_{\rm 12}}{F_{\rm tot}}$, the fraction of fluxes at two ends of the systems.
    \item $\frac{F_{\rm 12}}{F_{\rm 123}}$, the ratio between fluxes at two ends of the systems to fluxes outside the \HI{} disks.
    \item $\frac{R_{\rm 12}}{F_{\rm 123}}$, the ratio between residual flux at two ends of the systems to fluxes outside the \HI{} disks.
    \item $f^+_{\rm 123}$, the fraction of positive pixels outside \HI{} disks.
    \item $f^+_{\rm 12}$, the fraction of positive pixels at two ends of the systems.
\end{itemize}

In addition, we derive a disorder parameter $L$ describing the physical extent of \HI{}. It is calculated as:

\begin{equation} \label{equ:L}
    L \equiv \sqrt{L_{\rm R.A.} \times L_{\rm DEC.}} / (R_{\rm HI, prim} + R_{\rm HI, sec})
\end{equation}
where $L_{\rm R.A.}$ and $L_{\rm DEC.}$ are the numbers of pixels with \HI{} column densities higher than a uniform detection limit of $10^{18}$ cm$^{-2}$ in $R.A.$ and $DEC.$ directions, respectively. And $R_{\rm HI, prim}$ and $R_{\rm HI, sec}$ are the \HI{} disk sizes of the primary and secondary galaxies, respectively.

We also derive a disorder parameter $S$ which measures the clumpiness of \HI{} beyond the \HI{} disk. It is defined as the ratio of the sum of the absolute pixel values in the residual map and the sum of the pixel values in moment-0 map, with the pixels inside the two \HI{} disks excluded (Equation~\ref{equ:tidsig}).

\begin{equation} \label{equ:tidsig}
    S \equiv \frac{\sum_{\rm pix} |residual\,map|}{\sum_{\rm pix} moment-0\,map}
\end{equation}

All these parameters are viewed as measurements of disorder of the \HI{} morphologies and referred to as disorder parameters. And they are also measured from the controls. We further derive the difference between those of the interacting system and the median value of the controls. The difference is denoted as $\Delta (X)$, where $X$ is any disorder parameter described above. In particular, we refer to $\Delta (L)$ as the expansion of \HI{} and $\Delta (S)$ as the clumpiness of \HI{}.

\subsubsection{Gas surface density profile} \label{sec:SDprof}
We derive the gas surface density profiles of the interacting galaxies. Both \HI{} and H$_2$ profiles are measured from the corresponding moment-0 maps following a standard procedure of aperture photometry using annulus with increasing radius and fixed ellipticity and position angle determined in the optical. The same procedure is applied to the mocks and controls as well. We further derive the median profiles of controls for each interacting galaxy.

For \HI{} profiles we confirm that excluding pixels belonging to the other galaxy (Section~\ref{sec:HIdiv}) does not significantly change the results. And we note the systematic uncertainty introduced by inclination and spatial resolution which influence \HI{} observations more severely than those of H$_2$.

\subsubsection{Evaluation of SFR enhancement and HI excess} \label{sec:xcor}
We define the \HI{} excess ($\Delta \log M_{\rm HI}$) as follows:

\begin{equation} \label{equ:deltaMHI}
    \Delta \log M_{\rm HI} \equiv \log M_{\rm HI} - \log M_{\rm HI, MS}(M_*)
\end{equation}
where $M_{\rm HI, MS}(M_*)$ is the \HI{} mass predicted by HIMS.

\begin{figure*}
    \centering
    \includegraphics[width=0.45\textwidth]{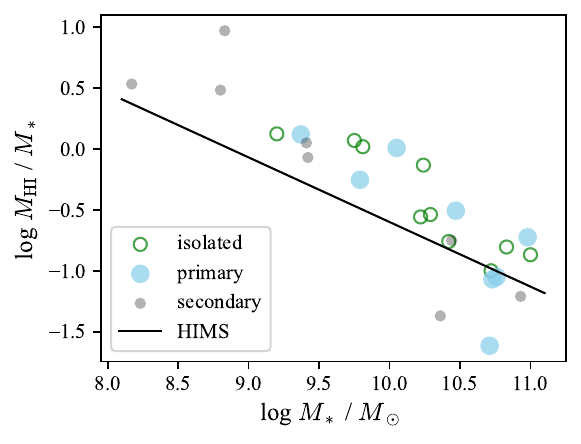}
    \includegraphics[width=0.455\textwidth]{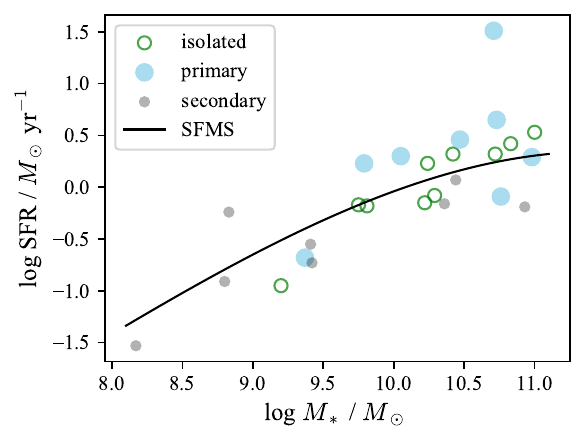}
\caption{FEASTS interacting and isolated galaxies on \HI{} fraction main sequence (HIMS, left panel) and star forming main sequence (SFMS, right panel). The primary galaxies are shown as large blue circles while the secondary galaxies are shown as small gray circles. The isolated galaxies are represented by open green circles. The corresponding main sequence are plotted as black curves. \label{fig:MS}}
\end{figure*}

The enhancement of SFR ($\Delta \log SFR$) is derived as Equation~\ref{equ:deltaSFR}.

\begin{equation} \label{equ:deltaSFR}
    \Delta \log SFR \equiv \log SFR - \log SFR_{\rm MS}(M_*)
\end{equation}
where $SFR_{\rm MS}(M_*)$ is the SFR predicted by SFMS.
We then refer to interacting galaxies with $\Delta \log SFR > 0$ as star formation enhanced galaxies. Interacting galaxies with $\Delta \log SFR < 0$ are referred to as star formation suppressed galaxies. We have six galaxies that are star formation enhanced and ten that are star formation suppressed. More than half of the primary galaxies are star formation enhanced and most of the secondary galaxies are star formation suppressed. The locus of the interacting and isolated galaxies on SFMS and HIMS are shown in Figure~\ref{fig:MS}.

In order to evaluate the change of gas reservoir caused by tidal interaction and take into account the effect of \HI{}-to-H$_2$ phase transition, we parameterize the excess of total neutral gas as the corrected \HI{} excess ($\Delta \log M_{\rm HI, cor}$, Equation~\ref{equ:mhicorr} and \ref{equ:mhicorr2}).

\begin{equation} \label{equ:mhicorr}
    \Delta M_{\rm HI, cor} = \Delta M_{\rm HI} + \Delta M_{\rm H_2}
\end{equation}
\begin{equation} \label{equ:mhicorr2}
    \Delta \log M_{\rm HI, cor} = \log [\frac{\Delta M_{\rm HI, cor}}{M_{\rm HI, MS}(M_*)} + 1]
\end{equation}
where $\Delta M_{\rm HI} = M_{\rm HI} - M_{\rm HI, MS}(M_*)$ and $\Delta M_{\rm H_2} = M_{\rm H_2} - M_{\rm H_2, MS}(M_*)$. $M_{\rm H_2, MS}(M_*)$ is the expected average H$_2$ mass at a given stellar mass \citep{2017ApJS..233...22S}. 
The excess of H$_2$ ($\Delta M_{\rm H_2}$) is estimated with $\Delta SFR$ based on the best-fit scaling relation between $\Delta \log M_{\rm H_2}$ and $\Delta \log SFR$ (Equation~\ref{equ:MH2SFR}), with data taken from the xCOLD GASS project \citep{2017ApJS..233...22S}.

\begin{equation} \label{equ:MH2SFR}
    \Delta \log M_{\rm H_2} = 0.515 \times \Delta \log SFR + 0.078
\end{equation}

A positive/negative gas mass excess is interpreted as the gas being over-abundant/deficient, thus being accreted/depleted. We note that estimating the H$_2$ excess and corrected \HI{} excess based on scaling relations basically assumes quasi-equilibrium systems, which is unlikely true for interacting and possibly starbursting galaxies. The estimation should thus only be taken as a first-order approximated, statistical view of the connection between \HI{}, H$_2$ and SFR.

\section{results} \label{sec:result}
\subsection{HI disorder caused by tidal interaction} \label{sec:distur}
\begin{figure*}
    \centering
    \includegraphics[width=0.43\textwidth]{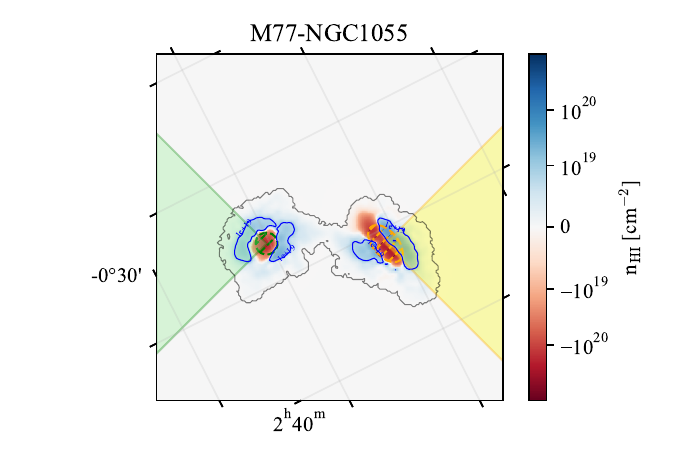}
    \includegraphics[width=0.43\textwidth]{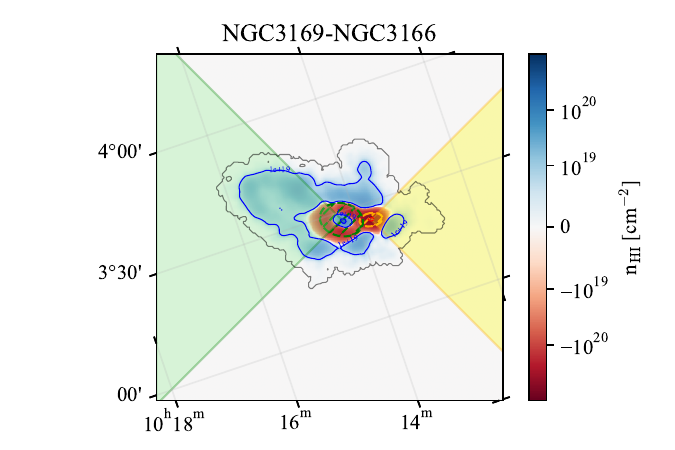}
    \includegraphics[width=0.43\textwidth]{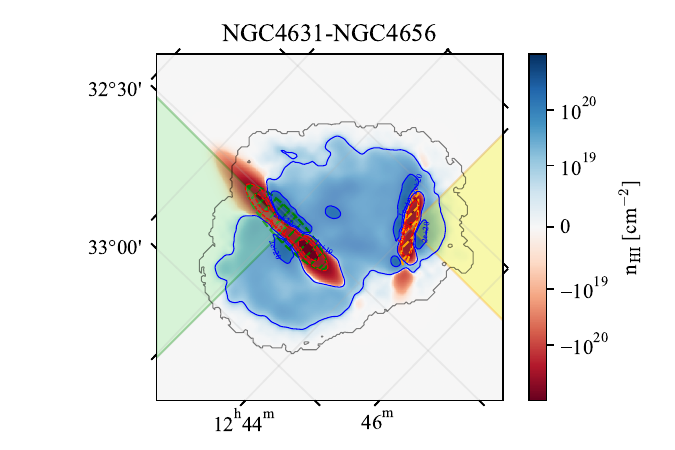}
    \includegraphics[width=0.43\textwidth]{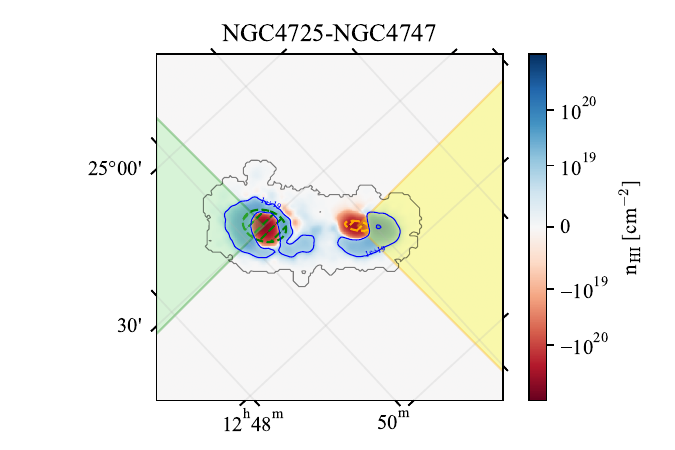}
    \includegraphics[width=0.43\textwidth]{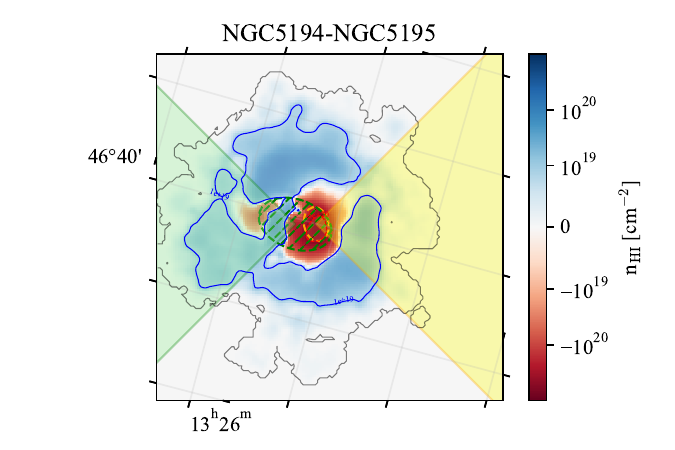}
    \includegraphics[width=0.43\textwidth]{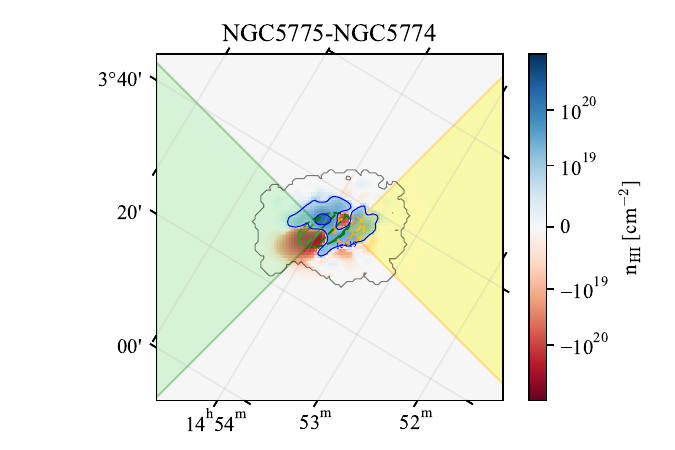}
    \includegraphics[width=0.43\textwidth]{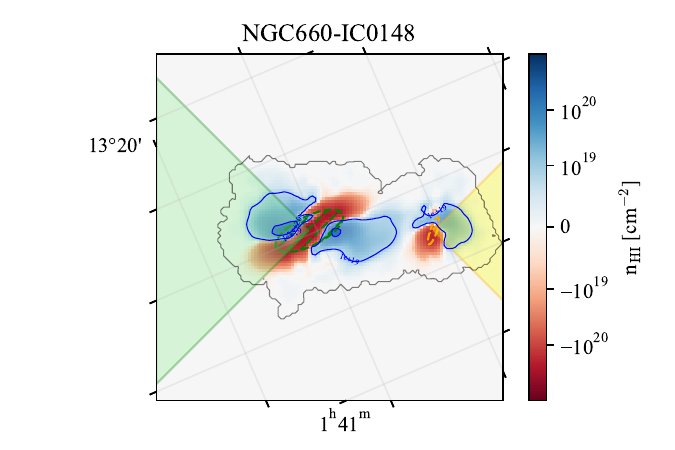}
    \includegraphics[width=0.43\textwidth]{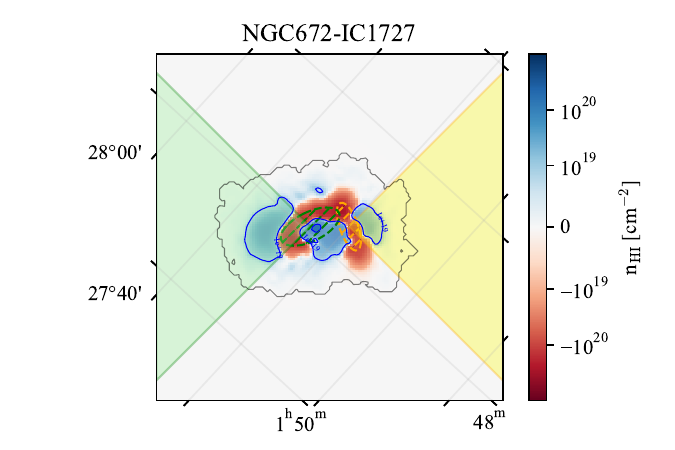}
\caption{The rotated residual maps of the interacting systems. The gray contour encloses pixels with column densities above $10^{18}$ cm$^{-2}$. Blue pixels indicate positive residuals, while red ones indicate negative residuals. The blue and red contours show the values of positive and negative residuals with labels, respectively. The green and orange hatched regions are \HI{} disks of primary and secondary galaxies. The regions at two ends of the system are shaded in green (connected to the primary galaxy) and yellow (connected to the secondary galaxy). \label{fig:res}}
\end{figure*}

In Figure~\ref{fig:res} we present the rotated residual maps. We see obvious negative regions (pixels in red) around the center of the galaxies in almost all interacting systems. And the pixels tend to be blue rather than red outside the \HI{} disks, especially at two ends of the system. While for the residual maps of the controls (Appendix~\ref{sec:resctrl}), the central regions are positive (pixels in blue) in most cases. The outskirts of the \HI{} disks are generally red, occasionally with blue pixels outside the hashed regions. Such differences beyond \HI{} disks are captured by the disorder parameters that we defined in Section~\ref{sec:disorder}.

\begin{figure*}
    \centering
    \includegraphics[width=0.65\textwidth]{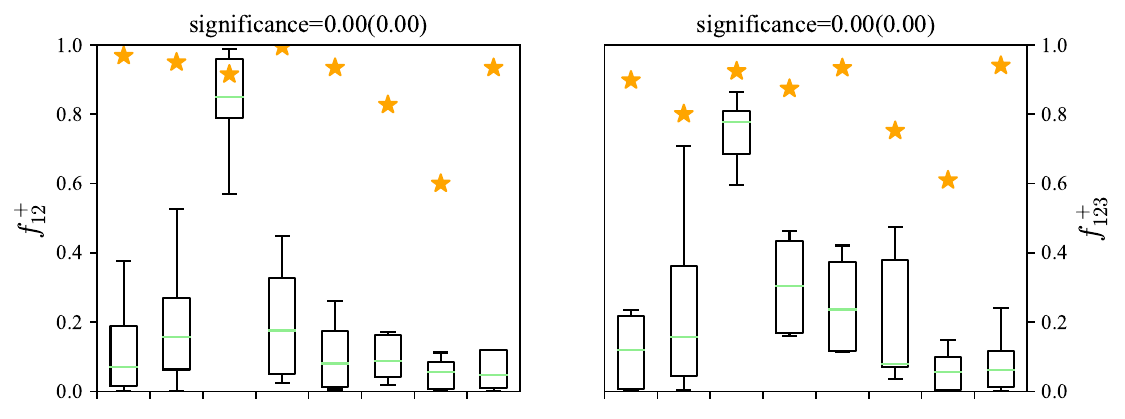}
    \includegraphics[width=0.65\textwidth]{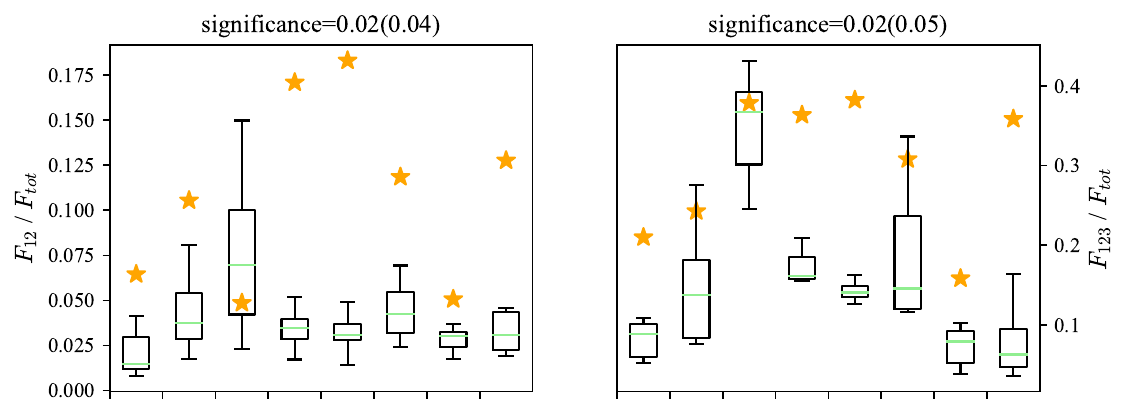}
    \includegraphics[width=0.65\textwidth]{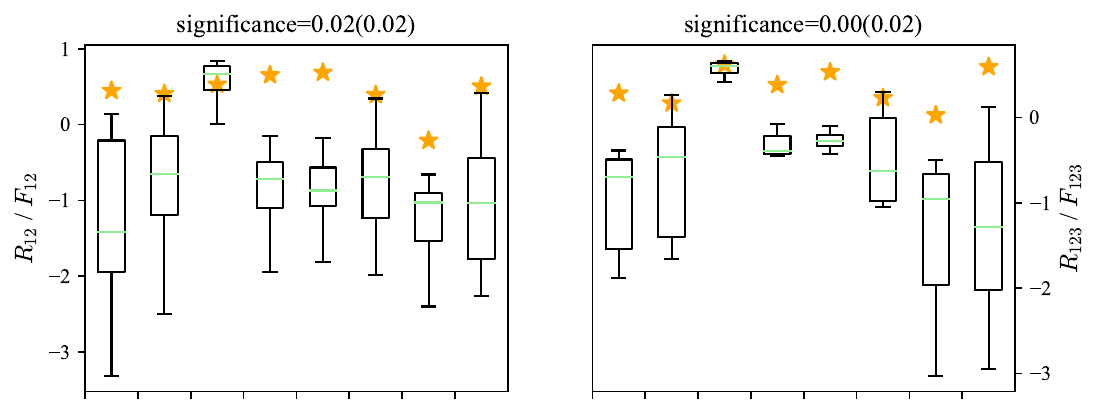}
    \includegraphics[width=0.65\textwidth]{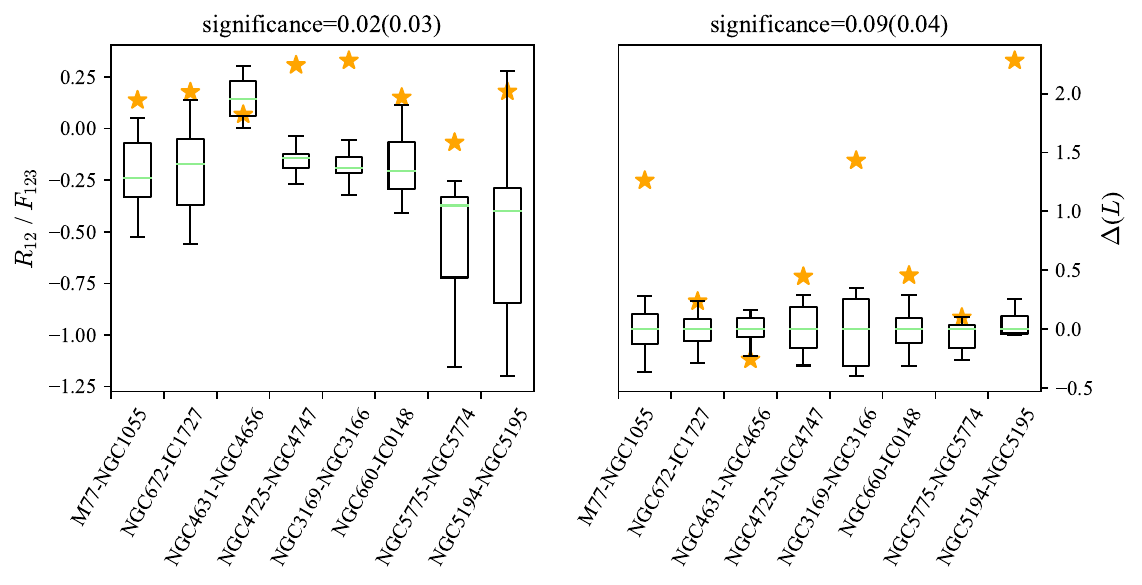}
\caption{The comparison between disorder parameters of the interacting systems and those of the controls. \textbf{Top row:} The fraction of positive pixels at two ends of the system ($f^+_{\rm 12}$) and outside the \HI{} disks ($f^+_{\rm 123}$). \textbf{Second row:}The fraction of fluxes at two ends of the system ($F_{\rm 12} / F_{\rm tot}$) and outside the \HI{} disks ($F_{\rm 123} / F_{\rm tot}$). \textbf{Third row:} The fraction of residual flux at two ends of the system ($R_{\rm 12} / F_{\rm 12}$) and outside the \HI{} disks ($R_{\rm 123} / F_{\rm 123}$). \textbf{Bottom row:} The ratio between residual flux at two ends of the systems to fluxes outside the \HI{} disks ($R_{\rm 12} / F_{\rm 123}$) and the expansion of \HI{} ($\Delta (L)$).
On the top of each figure, the significance of the difference is shown with the uncertainty in the following parenthesis. The orange stars represent the parameter values of the interacting systems. The error bars of disorder parameters are all well smaller than the symbols. The black boxes indicate the distribution of the disorder parameter derived from the controls. The box extends from $25-$ to $75-$percentile of the distribution with an green line at the median. The whiskers extend from the box by $1.5$ times the difference between $25-$ and $75-$percentile. \label{fig:resparam}}
\end{figure*}

We find significant difference in disorder parameters (Section~\ref{sec:disorder}) between the interacting systems and the controls. We use the K-S test to evaluate the significance of the difference in disorder parameter between interacting systems and controls. The uncertainty of the significance is estimated via bootstrap. In each bootstrap, we randomly extract one control for each interacting system and obtain a equal-size sample of disorder parameter for the controls. We then record the p-value of the K-S test between this extracted control sample and the interacting system sample. We repeat this procedure for $10,000$ times and obtain a distribution of p-value. We take the median and the standard deviation of this distribution as the significance of the difference and the corresponding uncertainty.
As shown in Figure~\ref{fig:resparam}, $f^+_{\rm 12}$, $f^+_{\rm 123}$, $\frac{F_{\rm 12}}{F_{\rm tot}}$, $\frac{F_{\rm 123}}{F_{\rm tot}}$, $\frac{R_{\rm 12}}{F_{\rm 12}}$, $\frac{R_{\rm 123}}{F_{\rm 123}}$, $\frac{R_{\rm 12}}{F_{\rm 123}}$ and $\Delta (L)$ are higher in interacting systems. Thus the disorder parameters successfully characterize disorder in \HI{} morphologies. The \HI{} in interacting systems preferentially piles outside the \HI{} disks, especially at two ends of the systems. And they tend to be more extended compared to those of the controls. On the other hand, the differences in $\frac{F_{\rm 12}}{F_{\rm 123}}$ and $\Delta (S)$ are not significant (significance $> 0.1$).

\begin{figure*}
    \centering
    \includegraphics[width=0.31\textwidth]{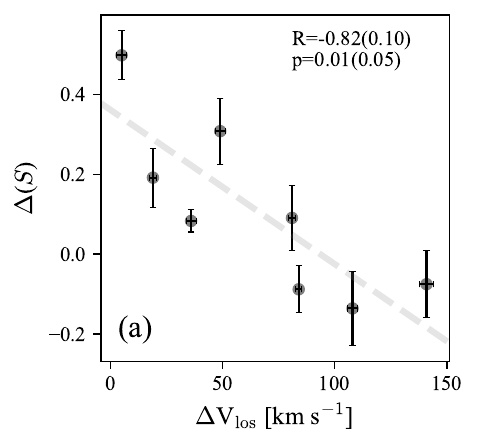}
    \includegraphics[width=0.3\textwidth]{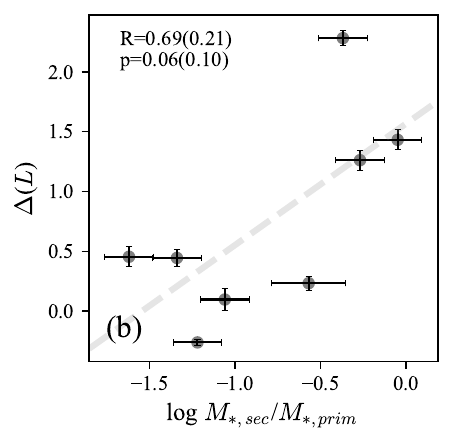}
\caption{Correlations between disorder parameters and properties of interaction.
\textbf{Panel (a):} The anti-correlation between $\Delta (S)$ and $\Delta V_{\rm los}$. \textbf{Panel (b):} The correlation between $\Delta(L)$ and $\log M_{\rm *, sec} / M_{\rm *, prim}$. The Pearson R coefficients and p-values are shown in the upper corners with the bootstrap uncertainties in the following parenthesis. The gray dashed lines are linear fit of the data points, for guiding the eye to the trend of the data. \label{fig:dependence}}
\end{figure*}

We further study the dependence of \HI{} morphological disorder on the configuration of tidal interaction by searching for correlations between disorder parameters and physical parameters directly related to tidal interaction.
The tidal interaction parameters include stellar mass ratios, projected distances and relative line-of-sight (l-o-s) velocities between the secondary and primary galaxies.
We measure the significance of the correlations between disorder parameters and galactic properties using the Pearson R coefficient and the corresponding p-value. The uncertainties are also obtained via bootstrap.
A summary of these statistics can be found in Table~\ref{tab:corr2}.
In Figure~\ref{fig:dependence} we present the most significant correlations, which are for the clumpiness and the expansion of \HI{}:
\begin{enumerate}
    \item $\Delta (S)$ (Equation~\ref{equ:tidsig}) anti-correlates with the relative l-o-s velocity $\Delta V_{\rm los}$. It is also significant if we normalize $\Delta V_{\rm los}$ by the maximum rotational velocity ($V_{\rm rot}$) of either primary or secondary galaxy.
    \item $\Delta(L)$ (Equation~\ref{equ:L}) correlates with the stellar mass ratio between secondary and primary galaxies ($\log M_{\rm *, sec} / M_{\rm *, prim}$).
\end{enumerate}

\begin{table*}
\caption{Significance of the correlations between \HI{} disorder and tidal interaction parameters. The Pearson R coefficient is shown for each correlation, with the corresponding p-value in the following parenthesis. The values for correlations with p-value $< 0.10$ are highlighted in bold. \label{tab:corr2}}
\begin{tabular}{c|rrr}
\hline
{} & {$\log M_{\rm *, sec} / M_{\rm *, prim}$} & {$\Delta V_{\rm los}$} & {$D_{\rm proj}$} \\
\hline
    $\Delta (\frac{F_{\rm 12}}{F_{\rm 123}})$ & 0.00(1.00) & 0.08(0.85) & 0.46(0.26) \\
    $\Delta (\frac{R_{\rm 12}}{F_{\rm 123}})$ & 0.47(0.24) & -0.14(0.75) & -0.21(0.62) \\
    $\Delta (\frac{F_{\rm 123}}{F_{\rm tot}})$ & 0.39(0.34) & -0.48(0.22) & -0.21(0.62) \\
    $\Delta (\frac{F_{\rm 12}}{F_{\rm tot}})$ & 0.32(0.44) & -0.37(0.36) & 0.01(0.99) \\
    $\Delta (\frac{R_{\rm 12}}{F_{\rm 12}})$ & 0.57(0.14) & 0.17(0.69) & 0.16(0.71) \\
    $\Delta (\frac{R_{\rm 123}}{F_{\rm 123}})$ & 0.38(0.35) & -0.13(0.76) & -0.39(0.35) \\
    $\Delta (f^+_{\rm 12})$ & 0.48(0.23) & 0.11(0.79) & -0.01(0.98) \\
    $\Delta (f^+_{\rm 123})$ & 0.53(0.18) & 0.14(0.75) & -0.18(0.66) \\
    $\Delta (S)$ & 0.25(0.55) & \textbf{-0.82(0.01)} & -0.21(0.62) \\
    $\Delta (L)$ & \textbf{0.69(0.06)} & -0.23(0.58) & -0.13(0.75) \\
    \hline
\end{tabular}
\end{table*}

These correlations show that both the relative velocities and the mass ratios between galaxies in the interacting systems strongly influence the \HI{} disorder. The first correlation shows that the clumpiness of \HI{} increases when the two galaxies are closer in the velocity domain. This correlation is also significant (p-value $< 0.05$) if Spearman or Kendall test is used. And the second correlation shows that the expansion of \HI{} increases when the two galaxies have more similar stellar masses. This correlation is not significant if Spearman or Kendall test is used.
Interestingly, we do not observe any significant correlations between \HI{} piling parameters ($\Delta (\frac{F_{\rm 12}}{F_{\rm 123}})$, $\Delta (\frac{R_{\rm 12}}{F_{\rm 123}})$, $\Delta (\frac{F_{\rm 123}}{F_{\rm tot}})$, $\Delta (\frac{F_{\rm 12}}{F_{\rm tot}})$, $\Delta (\frac{R_{\rm 12}}{F_{\rm 12}})$, $\Delta (\frac{R_{\rm 123}}{F_{\rm 123}})$, $\Delta (f^+_{\rm 12})$ and $\Delta (f^+_{\rm 123})$) and tidal interaction parameters. It implies that the piling of \HI{} outside \HI{} disks may depend more on other tidal interaction properties not quantified here, like the stage of interactions.

\subsection{Effects of tidal interaction on gas content and star formation rate} \label{sec:eff}
\begin{figure*}
    \centering
    \includegraphics[width=0.3\textwidth]{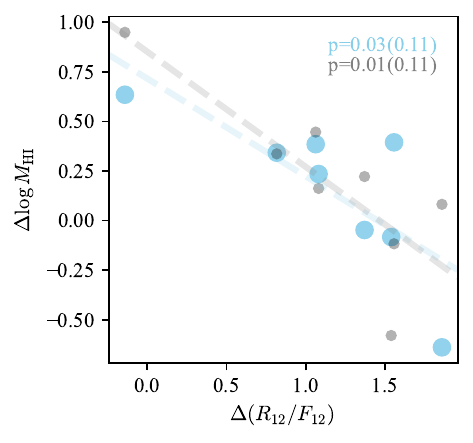}
    \includegraphics[width=0.31\textwidth]{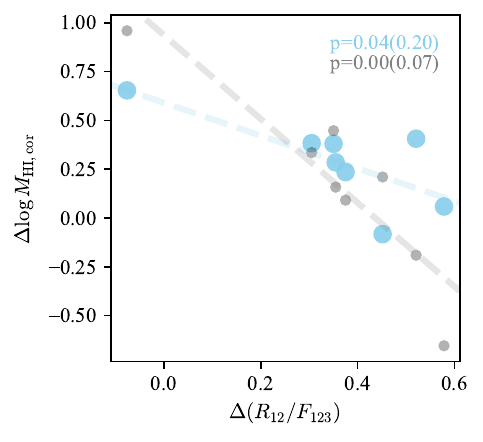}
    \includegraphics[width=0.3\textwidth]{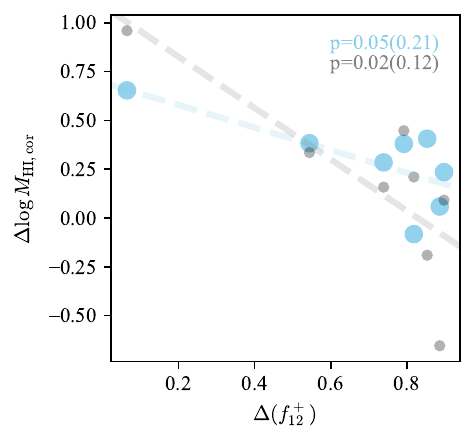}
\caption{The anti-correlations between (corrected) \HI{} excess and different disorder parameters. Larger blue circles represent primary galaxies and smaller gray ones represent secondary galaxies. The p-values of correlations for primary (blue) and secondary (gray) galaxies are shown in the upper right corners with the bootstrap uncertainties in the following parenthesis. The dashed lines in different colors are linear fit to the corresponding data points, for guiding the eye to the trend of the data.} \label{fig:effect}
\end{figure*}

We explore the effects of tidal interaction on the gas content and star formation rate of the galaxies by examining correlations between gas excess, SFR excess and the disorder parameters. In Table~\ref{tab:corr} we present the Pearson R coefficients and p-value for all the combinations of correlations, for primary galaxies and secondary galaxies, respectively.
For primary galaxies, there are significant (p-value $< 0.05$) anti-correlations for the \HI{} excess ($\Delta \log M_{\rm HI}$) with $\Delta (\frac{R_{\rm 12}}{F_{\rm 12}})$. The corrected \HI{} excess ($\Delta \log M_{\rm HI, cor}$) anti-correlates with $\Delta(\frac{R_{\rm 12}}{F_{\rm 123}})$ and $\Delta(f_{\rm 12}^+)$. These significant correlations are shown in Figure~\ref{fig:effect}, highlighting the connection between \HI{} dragged outside \HI{} disks onto the two ends of the system and the deficiency in \HI{} and total neutral gas for primary galaxies. However, the p-value of the correlations have large uncertainties, reflecting the trends to be primarily driven by a few galaxies and caution on the significance of the correlations. We also note that these correlations for primary galaxies are not significant based on Spearman or Kendall tests (Appendix~\ref{sec:spke}). And we see significant anti-correlation between star formation rate enhancement and $\Delta (\frac{F_{\rm 12}}{F_{\rm 123}})$ if Spearman test is used.

For secondary galaxies, both \HI{} excess and corrected \HI{} excess significantly anti-correlate with $\Delta(\frac{R_{\rm 12}}{F_{\rm 123}})$, $\Delta(\frac{F_{\rm 123}}{F_{\rm tot}})$, $\Delta(\frac{R_{\rm 12}}{F_{\rm 12}})$, $\Delta(\frac{R_{\rm 123}}{F_{\rm 123}})$, $\Delta(f_{\rm 12}^+)$, $\Delta(f_{\rm 123}^+)$ and $\Delta (L)$. They are strongly related to the piling of \HI{} outside disks and \HI{} expansion, but not with the \HI{} clumpiness. The star formation rate enhancement also show significant anti-correlation with $\Delta(\frac{R_{\rm 12}}{F_{\rm 123}})$, $\Delta(\frac{F_{\rm 123}}{F_{\rm tot}})$, $\Delta(\frac{F_{\rm 12}}{F_{\rm tot}})$, $\Delta(\frac{R_{\rm 12}}{F_{\rm 12}})$, $\Delta(f_{\rm 12}^+)$ and $\Delta(f_{\rm 123}^+)$. They are mostly linked with the piling of \HI{} outside disks. The \HI{} piling outside \HI{} disks is a sensitive probe for change in \HI{} abundance and star formation levels for secondary galaxies. Using Spearman or Kendall tests generally only produces slightly different p-values for \HI{} excess and corrected \HI{} excess correlations (Appendix~\ref{sec:spke}). Except for that $\Delta(\frac{R_{\rm 123}}{F_{\rm 123}})$ is not significantly correlated with either \HI{} excess or corrected \HI{} excess. Nearly half of the disorder parameters that originally show significant Pearson correlations become insignificantly correlated with SFR enhancements when using Spearman or Kendall tests. The different levels of robustness suggest that \HI{} abundances respond more closely to \HI{} disorder levels than the star formation rates.

\begin{table*}
\caption{Significance of the correlations between gas content, star formation and \HI{} disorder. The Pearson R coefficient is shown for each correlation, with the corresponding p-value in the following parenthesis. The values for the significant correlations (p-value $< 0.05$) are highlighted in bold. Correlations for primary galaxies are presented on the left while those for secondary galaxies are on the right. \label{tab:corr}}
\begin{tabular}{c|rrr|rrr}
\hline
{} & \multicolumn{3}{c}{primary galaxies} & \multicolumn{3}{c}{secondary galaxies}\\
{} & {$\Delta \log M_{\rm HI}$} & {$\Delta \log M_{\rm HI, cor}$} & {$\Delta \log SFR$} & {$\Delta \log M_{\rm HI}$} & {$\Delta \log M_{\rm HI, cor}$} & {$\Delta \log SFR$} \\
\hline
    $\Delta (\frac{F_{\rm 12}}{F_{\rm 123}})$ & -0.07(0.86) & -0.24(0.56) & -0.40(0.32) & 0.03(0.95) & 0.02(0.95) & -0.51(0.20) \\
    $\Delta (\frac{R_{\rm 12}}{F_{\rm 123}})$ & -0.48(0.23) & \textbf{-0.74(0.04)} & -0.10(0.81) & \textbf{-0.91(0.00)} & \textbf{-0.91(0.00)} & \textbf{-0.93(0.00)} \\
    $\Delta (\frac{F_{\rm 123}}{F_{\rm tot}})$ & -0.35(0.40) & -0.70(0.05) & -0.15(0.72) & \textbf{-0.92(0.00)} & \textbf{-0.93(0.00)} & \textbf{-0.80(0.02)} \\
    $\Delta (\frac{F_{\rm 12}}{F_{\rm tot}})$ & -0.23(0.58) & -0.64(0.09) & -0.41(0.31) & -0.67(0.07) & -0.68(0.06) & \textbf{-0.88(0.00)} \\
    $\Delta (\frac{R_{\rm 12}}{F_{\rm 12}})$ & \textbf{-0.76(0.03)} & -0.69(0.06) & 0.22(0.60) & \textbf{-0.81(0.01)} & \textbf{-0.80(0.02)} & \textbf{-0.80(0.02)} \\
    $\Delta (\frac{R_{\rm 123}}{F_{\rm 123}})$ & -0.54(0.17) & -0.66(0.08) & 0.23(0.59) & \textbf{-0.92(0.00)} & \textbf{-0.92(0.00)} & -0.62(0.10) \\
    $\Delta (f^+_{\rm 12})$ & -0.64(0.09) & \textbf{-0.72(0.05)} & 0.03(0.95) & \textbf{-0.80(0.02)} & \textbf{-0.78(0.02)} & \textbf{-0.86(0.01)} \\
    $\Delta (f^+_{\rm 123})$ & -0.64(0.09) & -0.63(0.09) & 0.22(0.60) & \textbf{-0.90(0.00)} & \textbf{-0.89(0.00)} & \textbf{-0.78(0.02)} \\
    $\Delta (S)$ & -0.03(0.94) & -0.40(0.33) & -0.18(0.67) & -0.66(0.08) & -0.68(0.06) & -0.35(0.39) \\
    $\Delta (L)$ & -0.55(0.16) & -0.50(0.21) & 0.30(0.47) & \textbf{-0.94(0.00)} & \textbf{-0.94(0.00)} & -0.55(0.16) \\
    \hline
\end{tabular}
\end{table*}

\subsection{Gas distribution in interacting systems}
\begin{figure}
    \centering
    \includegraphics[width=0.47\textwidth]{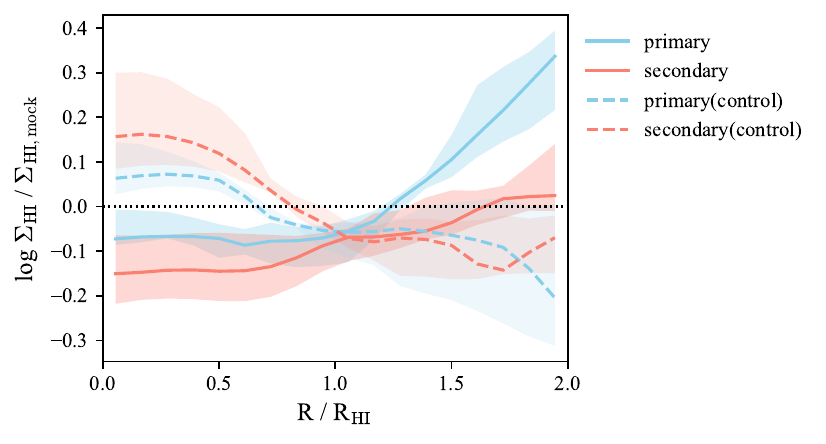}
\caption{The median profiles of \HI{} surface density ratio. Solid lines are used for interacting galaxies and dashed lines for the controls. The profiles of primary and secondary galaxies are plotted in blue and red, respectively. The shaded regions represent the uncertainties ($15.9$ and $84.1$ percentile) of the corresponding profile. \label{fig:profHI}}
\end{figure}

In Figure~\ref{fig:profHI} we present the median \HI{} surface density profile of the interacting galaxies and the controls. The median \HI{} profiles of primary and secondary galaxies are separately shown. The y-axis represents the surface density ratio between the interacting galaxies and the mock ($\log \Sigma_{\rm HI} / \Sigma_{\rm HI, mock}$), so that the difference between a real interacting galaxy and an idealized mock is also highlighted.

We find that firstly, between 0.75 and 1.25 $R_{\rm HI}$, both interacting galaxies and control galaxies show negative offsets of $\sim 0.1$ dex from the mock galaxies, reflecting possible systematic bias when constructing the mock galaxies based on previous interferometric measurements, or the relatively small control sample size. But at smaller and larger radius, the interacting and control galaxies show contrary offsets with respect to the mock galaxies, reflecting real differences between these two types of galaxies. While the controls of both primary and secondary galaxies have significantly higher \HI{} surface density than the mocks at $R / R_{\rm HI} \lesssim 0.75$, interacting galaxies show significant suppression of \HI{} surface density.
And the \HI{} surface density of both primary and secondary galaxy are significantly higher than those of the controls at $R / R_{\rm HI} \gtrsim 1.25 - 1.5$. This suggests that interacting galaxies have less \HI{} inside and more outside their \HI{} disks.

\begin{figure*}
    \centering
    \includegraphics[width=0.4\textwidth]{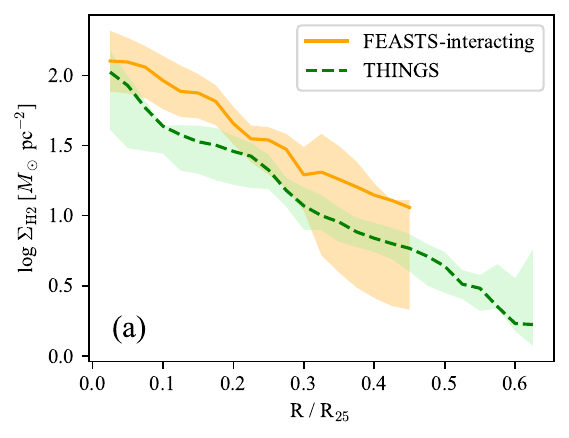}
    \includegraphics[width=0.4\textwidth]{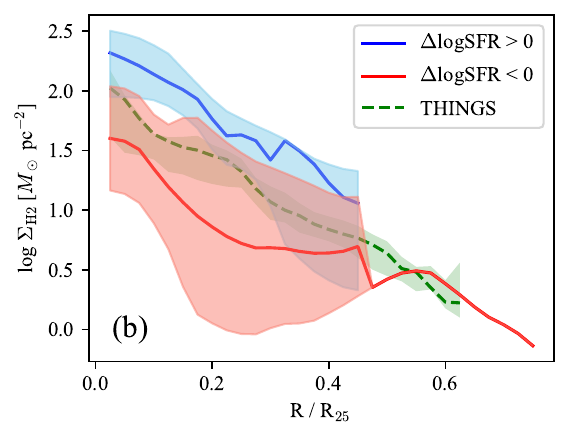}
    \includegraphics[width=0.4\textwidth]{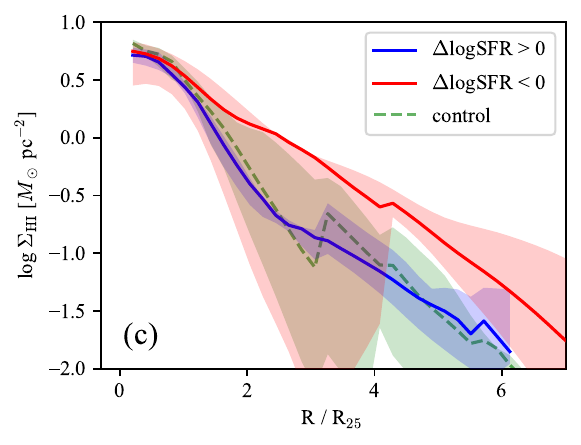}
     \includegraphics[width=0.4\textwidth]{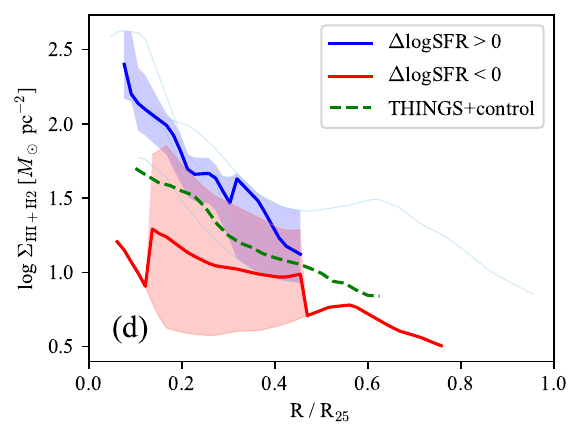}
\caption{The gas surface density profiles of the primary galaxies and the corresponding controls. \textbf{Panel (a):} The median H$_2$ surface density profiles compared to THINGS galaxies \citep{2008AJ....136.2782L}. \textbf{Panel (b):} The median H$_2$ surface density profiles of the star formation enhanced and suppressed galaxies compared to THINGS galaxies. \textbf{Panel (c):} The median \HI{} surface density profiles of the star formation enhanced and suppressed galaxies compared to FEASTS isolated galaxies (Section~\ref{sec:ctrl}). \textbf{Panel (d):} The median (in dark color) and individual (in light color) profiles of total neutral gas (\HI{} $+$ H$_2$) surface density compared to a virtual control produced by adding up the median \HI{} profile of FEASTS isolated galaxies and H$_2$ profile of THINGS galaxies. Profiles of all, star formation enhanced and star formation suppressed primary galaxies are shown in orange, blue and red, respectively. The profiles of the corresponding controls are shown as green dashed curves. The shaded regions represent the uncertainties ($15.9$ and $84.1$ percentile) of the profiles in the corresponding color. \label{fig:prof-SF-prim}}
\end{figure*}

These trends of gas surface density are further studied in Figure~\ref{fig:prof-SF-prim} for primary galaxies with enhanced and suppressed integral SFR. We can see that the star formation suppressed primary galaxies have higher \HI{} surface densities outside the \HI{} disks. The H$_2$ surface density profiles are also presented for a subset of our sample for which CO image observations are available. We can see that the primary galaxies have higher H$_2$ gas surface densities than typical star-forming galaxies (from THINGS, \citealt{2008AJ....136.2782L}), which is mainly driven by the star formation enhanced ones.

\section{discussion} \label{sec:dis}
\subsection{Quantified relation between tidal strength and perturbed HI morphologies} \label{sec:HImorph}
Tidal interactions efficiently perturb and re-distribute \HI{} of galaxies. \HI{} tidal structures have frequently been found in previous studies, mostly based on interferometric \HI{} images \citep[e.g.,][]{1996AJ....111..655H}. We have quantitatively characterized tidally perturbed morphologies using a new set of morphological parameters (Figure~\ref{fig:resparam}), with new \HI{} images that do not miss extended and low-column density \HI{}, unlike the case for interferometric data (Appendix~\ref{sec:AAGMRT}).

We have shown with radial profiles in comparison to control galaxies (Figure~\ref{fig:profHI}) that the \HI{} surface densities at large radius are enhanced for both primary and secondary galaxies. This is also indicated by the systematically larger $\Delta (L)$ parameters of the interacting sample in comparison to the control sample (Figure~\ref{fig:resparam}).
A series of parameters quantifying the disorder of \HI{} morphologies show that the interacting systems have systematically higher disorder in \HI{} than the controls (Figure~\ref{fig:resparam}).
We further showed in Figure~\ref{fig:dependence} the dependence of $\Delta (S)$ and $\Delta (L)$ on relative velocity \citep[e.g.,][]{2000ApJ...530..660B, 2008ApJ...683...94O} and mass ratio \citep[e.g.,][]{2008AJ....135.1877E, 2018MNRAS.479.3381B}. It is likely that the strength of tidal interaction, which by theory is related to the galaxy mass ratio and orbits \citep[e.g.,][]{1972ApJ...178..623T}, has determined the significance of \HI{} tidal structures such as bridges and tails \citep[e.g.,][]{2004IAUS..217..390M}.

These results indicate that interacting systems have systematically larger expansion of \HI{} than unperturbed systems, and the morphological parameters successfully characterize the significance of tidal perturbations.

We notice that the NGC4631-NGC4656 system is a significant outlier, for which the disorder parameters tend to be lower than the control galaxies in many cases (Figure~\ref{fig:resparam}). This is surprising, as most previous studies conclude this system as having experienced significant tidal perturbations \citep[e.g.,][]{1978A&A....65...37W, 1994A&A...285..833R}. It is possibly because NGC4631-NGC4656 is the only interacting pair having both galaxies extremely inclined ($incl. > 80^\circ$). Moreover, most of the tidally perturbed \HI{} lies between the two galaxies, which is likely due to the specific interacting orbits \citep{1978A&A....65...47C}, while most of the disorder parameters (e.g., $f^+_{\rm 12}$, $\frac{F_{\rm 12}}{F_{\rm tot}}$, $\frac{R_{\rm 12}}{F_{\rm 12}}$ and $\frac{R_{\rm 12}}{F_{\rm 123}}$) are biased toward detecting signals beyond the line linking the pair galaxies. As a result, the disorder parameters may be insensitive to the perturbation levels exhibited in systems like the NGC4631-NGC4656 pair. We leave such a caveat to future studies when more samples are available.

\subsection{Do tidal interaction always induce enhanced SFR?}
Theoretically, two basic neutral gas properties influence the SFR in galaxies --- the distribution and the kinematics, which determine the amount of available material at star-forming regions, and whether star formation can efficiently happen with material on site, respectively \citep{2008AJ....136.2782L, 2008AJ....136.2846B, 2019A&A...622A..64B}. The tidal interaction influences both properties, by stripping the gas from both primary and secondary galaxies, and by compressing the gas and driving inflows \citep{2022ApJ...927...66W, 2012MNRAS.426..549S}.

It is clear that these \HI{}-rich interacting galaxies are not always enhanced in their SFR (Section~\ref{sec:xcor}). In fact, almost all of the secondary galaxies have suppressed SFR. It seems that gas removal due to tidal stripping is dominantly more important than gas compression due to tidal shocks and ram pressure in these relatively low-mass galaxies.
For primary galaxies, there is no significant correlation between the star formation rate enhancement and any \HI{} disorder based on Pearson tests (Table~\ref{tab:corr}). The majority of them (six out of eight) being star formation enhanced suggests that the possible dominant mechanism could be tidally induced gas compression and inflow in the inner disk increasing the inner-disk H$_2$ and total neutral gas surface densities (Figure~\ref{fig:prof-SF-prim}). On the other hand, the significant anti-correlation between star formation rate enhancement and $\Delta (\frac{F_{\rm 12}}{F_{\rm 123}})$ based on Spearman test (Appendix~\ref{sec:spke}) suggests that the effects of gas removal due to tidal stripping may be non-negligible. The configuration and stage of interaction might also play a role in determining $\Delta \log SFR$ of the primary galaxies.
These results demonstrate that, a statistical sample of spatially resolved gas images, and analysis of the gas distribution and kinematics are crucial for us to physically understand the suppression and enhancement in tidally interacting galaxies.

\subsection{Does the tidal interaction deplete or accrete HI gas?}
Table~\ref{tab:corr} show strong and/or significant anti-correlation between corrected \HI{} excess and almost all \HI{} disorder parameters for secondary galaxies, strongly supporting the relation between \HI{} stripping and \HI{} depletion in these minor galaxies in interactions.
The \HI{} excess or corrected \HI{} excess also anti-correlates with several \HI{} disorder parameters for primary galaxies based on Pearson test, and the simplest interpretation is that stronger tidal interactions also tend to deplete the \HI{} in them. But the situation can be more complex than that for secondary galaxies.

Firstly, there is no significant correlation between (corrected) \HI{} excess and any disorder parameter for primary galaxies based on Spearman or Kendall tests (Appendix~\ref{sec:spke}). Secondly, the significant Pearson correlations between \HI{} disorders and corrected \HI{} excess of primary galaxies are largely driven by the outlier NGC4631 (Figure~\ref{fig:effect}). As discussed in Section~\ref{sec:HImorph}, possibly due to a combined effect of inclination and encounter orbit, the disorder parameters may be insensitive to the perturbation level in this system. If we remove this galaxy from the sample, the correlation significance between $\Delta(\frac{R_{\rm 12}}{F_{\rm 123}})$,  $\Delta(f_{\rm 12}^+)$ and corrected \HI{} excess drops to $0.30$ and $0.38$, respectively. This is also indicated by the large uncertainty in p-value and non-correlations based on Spearman or Kendall tests (Appendix~\ref{sec:spke}). If the total neutral gas is actually independent from \HI{} disorder, then a balance between the enhanced gas consumption (e.g., star formation) and increased rate of gas replenishment (e.g., accretion from CGM) is indicated. There are theoretical predictions that, CGM efficiently cools during tidal interaction, which happens at the significantly increased interface between \HI{} and CGM through turbulence mixing and thermal conduction \citep{2022MNRAS.509.2720S}. The theory is supported in several detailed case studies \citep[][Lin et al. in prep.]{2023ApJ...944..102W}.

Thirdly, gas accretion may happen at interaction stages not captured by those significantly correlated \HI{} disorder parameters for primary galaxies.
Statistical observational studies have been conducted in the literature, comparing mergers at different stages with control samples. They found \HI{} enhancement in the post-merger sample \citep{2015MNRAS.448..221E}, and do not always find \HI{} deficiency in the interacting sample \citep{2018ApJS..237....2Z, 2019ApJ...870..104S, 2022ApJ...934..114Y}. The distinct fates of \HI{} in the literature studies imply that the tidal depletion, star formation consumption, CGM cooling, \HI{} inflow may dominate at different stages.
In Table~\ref{tab:corr}, the significant anti-correlations for the (corrected) \HI{} excess are with the parameters describing the mass of \HI{} being dragged outside \HI{} disks onto the two ends ($\Delta(\frac{R_{\rm 12}}{F_{\rm 123}})$ and  $\Delta(f_{\rm 12}^+)$ ), but not those describing the extents or clumpiness of \HI{} ($\Delta (L)$ and $\Delta (S)$). A possible interpretation is thus that, \HI{} is firstly depleted when a considerable fraction of \HI{} is dragged out of \HI{} disks, corresponding to the correlation between piling of \HI{} at two ends of the system and \HI{} deficiency. But later the situation changes when the perturbed \HI{} expands to large areas, enlarging the interface with the CGM and boosting the cooling \citep{2024ApJ...968...48W, 2022MNRAS.509.2720S}.
If it was a one-way process for \HI{} to be depleted after \HI{} is dragged out \HI{} disks, then a simple picture expected would be the \HI{} becoming more deficient while expanding (i.e., next step of piling only at two ends of the system), and the morphology becoming more clumpy while \HI{} evaporates and/or gets ionized. Neither trends are observed.

Besides, the causality between \HI{} deficiency and piling of \HI{} beyond \HI{} disks may be the other way round, such that initially \HI{} deficient systems tend to produce higher fraction of tidally stripped \HI{} staying outside the initially unperturbed \HI{} disk. This is possible because \HI{}-poor galaxies having less extended \HI{} disks tend to have weaker hydrodynamic effects (e.g. shocks and ram pressure) during interaction \citep{2019ApJ...882...14M}, and thus less inflow.

In summary, high level of \HI{} (and neutral gas) deficiency seems to be associated with the mass fraction of \HI{} stripping, while \HI{} accretion may just start or be delayed to later stages when \HI{} expands to large areas. Finally, we note that systematic uncertainties in separating \HI{} into primary and secondary galaxies may affect the results.

\section{summary and conclusion} \label{sec:con}
In this work we investigate the fate of gas and star formation in interacting galaxies. Thanks to the FEASTS survey, we are able to fully map \HI{} gas in the interacting systems down to column density of $\sim 10^{18}$ cm$^{-2}$ and physical scale up to $\sim 100$ kpc. By building mocks and controls, we quantitatively describe disorder in \HI{} morphology using a set of new morphological parameters. We further explore the effects of tidal interaction on the properties of \HI{}, H$_2$ and the star formation rate of the interacting galaxies. Possible mechanisms that are responsible for influencing the gas reservoir and \HI{} content, and those for causing star formation enhancement or suppression, are discussed. The key conclusions are as follows:

{\begin{enumerate}
    \item     Compared to a control sample, \HI{} in interacting systems is distributed over a larger radius, has larger expansion, and preferentially accumulates at the opposing sides of the system. Such \HI{} disorder is well characterized by the disorder parameters we defined. They show strong dependence on tidal interaction parameters stellar mass ratio and relative velocities between primary and secondary galaxies.
    \item The generality of star formation suppression in the secondary galaxies points to the dominant effect of gas removal over induced compression and inflow during tidal interaction. While for the primary galaxies, star formation is more likely to be enhanced and independent from \HI{} disorder.
    \item The net consequence of a tidal interaction on the gas reservoir of the galaxies is a significant decrease in \HI{} content and total neutral gas content for the secondary galaxies, while for the primary galaxies the relation is less strong or significant. These correlations strongly supporting the relation between tidal stripping and gas depletion in the secondary galaxies, and suggest a more complex situation for the primary galaxies.
\end{enumerate}}

The complex situation of the primary galaxies may relate to the different dominant effects at different stages of the interaction. Gas can firstly be depleted when it is dragged outside the \HI{} disks, and later be enriched when accretion from CGM becomes dominant. Such speculation also partly explains the absence of significant correlation between star formation rate enhancement and \HI{} disorder.

We note that the conclusions we reach in this work are largely based on a small sample of \HI{}-rich galaxies. The dynamical range of the properties is rather limited. Whether the physical picture of gas depletion, accretion and star formation is valid in more general situations needs further testing.
A more complete picture of the interplay between CGM, ISM and galaxy evolution in interacting systems will be further and better understood through combining with other multi-wavelength data and high-resolution hydrodynamic simulations.

\section*{Acknowledgements}
We thank the anonymous referee for constructive comments.
This work made use of the data from FAST (Five-hundred-meter Aperture Spherical radio Telescope)\footnote{https://cstr.cn/31116.02.FAST}. FAST is a Chinese national mega-science facility, operated by National Astronomical Observatories, Chinese Academy of Sciences.
JW thanks support of research grants from  Ministry of Science and Technology of the People's Republic of China (NO. 2022YFA1602902), National Science Foundation of China (NO. 12073002, 12233001, 8200906879), and the China Manned Space Project.
SW thanks Dr. Chen and Luna.

\section*{Data Availability}
\HI{} data from FEASTS will be published online\footnote{https://github.com/FEASTS/LVgal/wiki}. All other data supporting this study is available from the corresponding author upon reasonable request.



\bibliographystyle{mnras}
\bibliography{mnras_example} 




\appendix
\section{channel and moment-0 maps compared to previous observations} \label{sec:AAGMRT}
In Figure~\ref{fig:chan3169} we compare the channel maps of the NGC3169-NGC3166 system from the FEASTS data cube and those from the ALFALFA data cube. Similarly the channel maps of the NGC4725-NGC4747 system from FEASTS and ALFALFA are presented in Figure~\ref{fig:chan4725}.
The FEASTS channel maps are interpolated to match the velocities of the ALFALFA data cube. The field-of-view and color scales are identical between different data of the same system.
The rms level of FEASTS data cube is significantly lower than that of the ALFALFA data cube. Thus FEASTS data cube reveals more faint structures than ALFALFA. The spatial resolution of FEASTS is also better, showing significantly more details of the \HI{} morphology across different velocity slices.

\begin{figure*}
    \centering
    \includegraphics[width=0.49\textwidth]{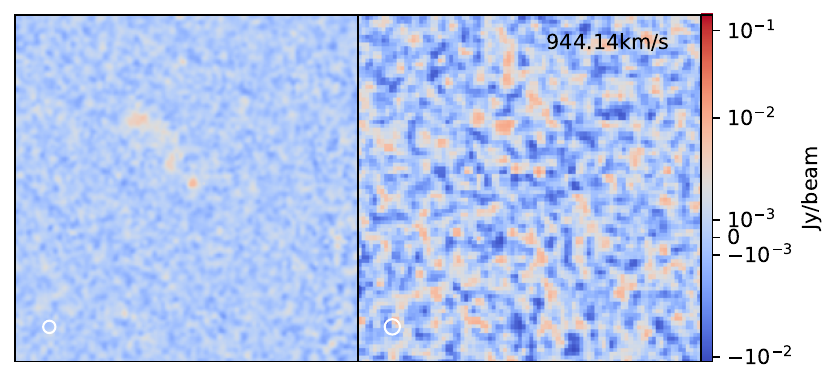}
    \includegraphics[width=0.49\textwidth]{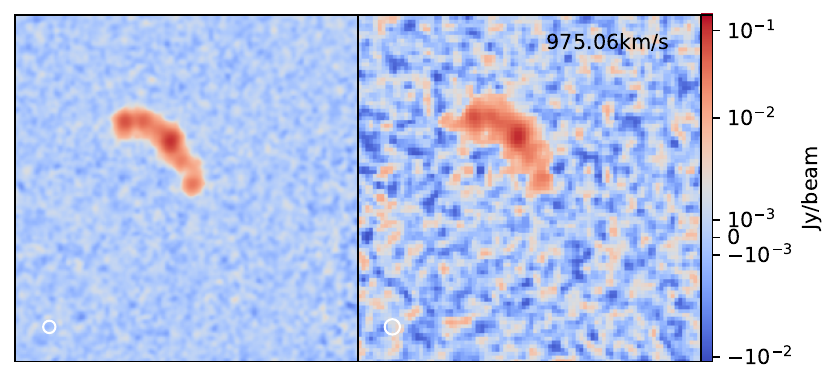}
    \includegraphics[width=0.49\textwidth]{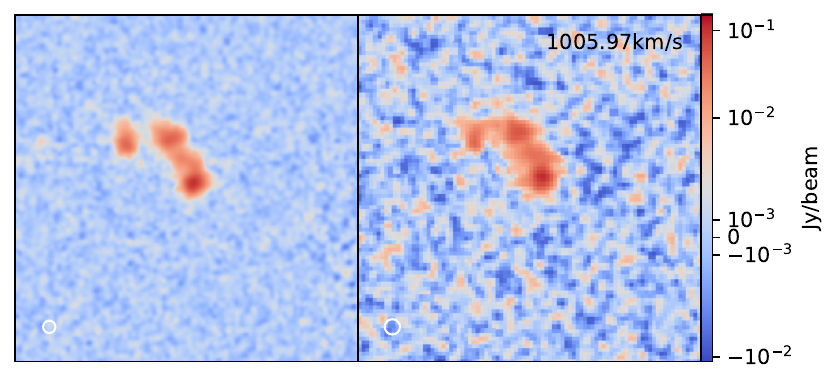}
    \includegraphics[width=0.49\textwidth]{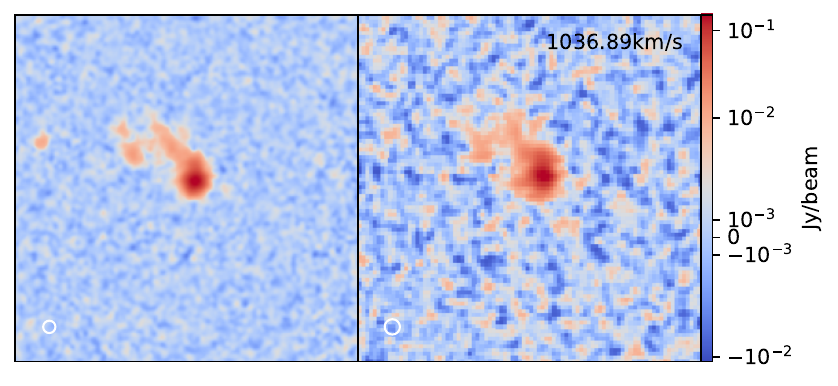}
    \includegraphics[width=0.49\textwidth]{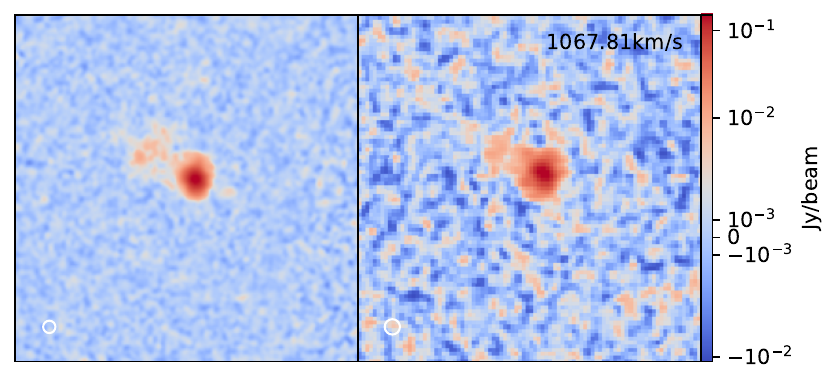}
    \includegraphics[width=0.49\textwidth]{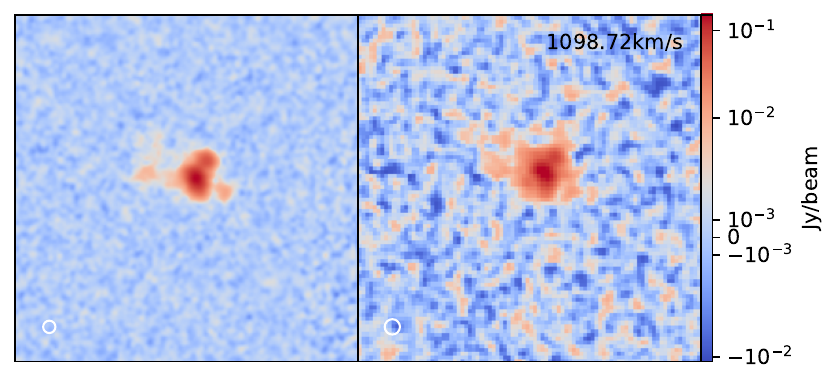}
    \includegraphics[width=0.49\textwidth]{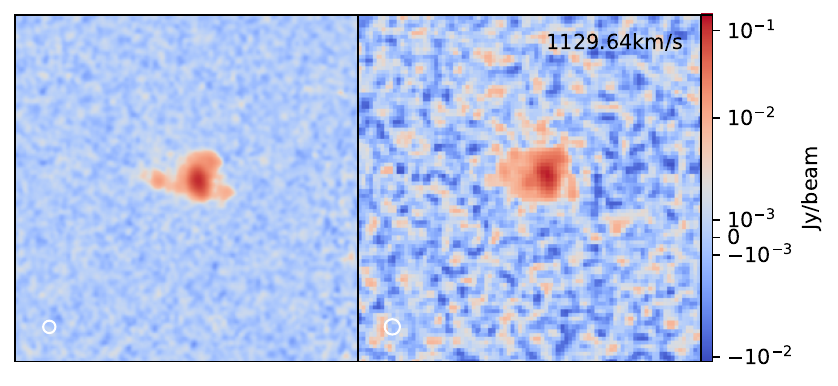}
    \includegraphics[width=0.49\textwidth]{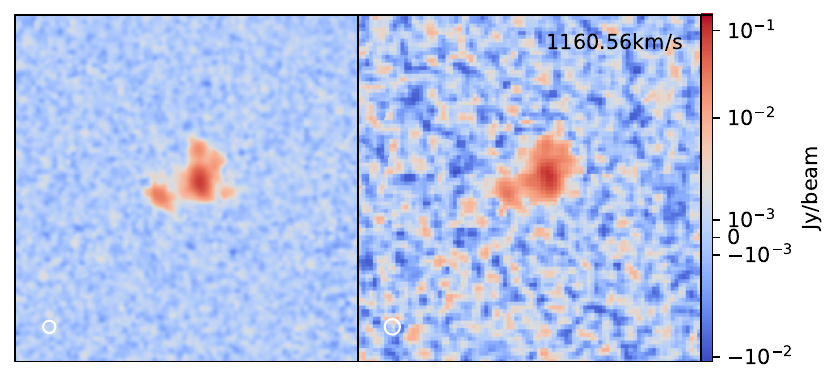}
    \includegraphics[width=0.49\textwidth]{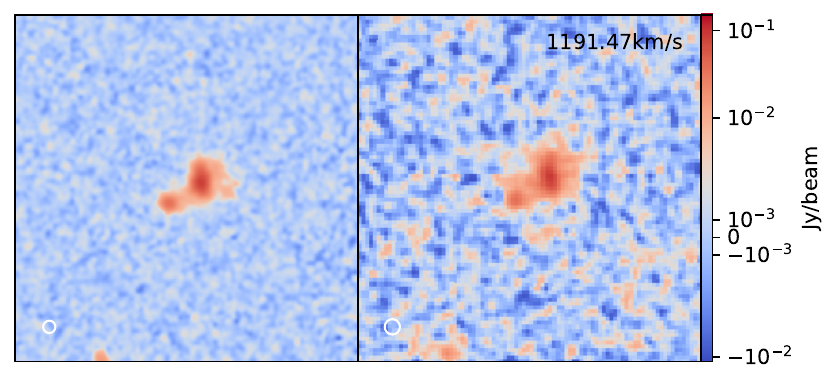}
    \includegraphics[width=0.49\textwidth]{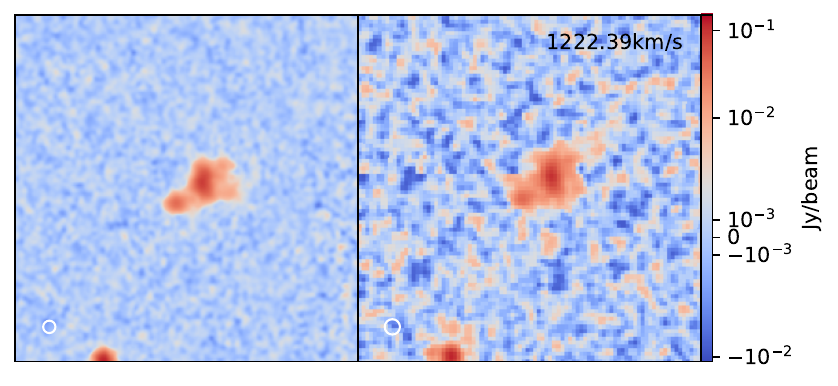}
\caption{The channel maps of the NGC3169-NGC3166 system from FEASTS (left panels) and ALFALFA (right panels). The velocity of the channel is shown in the upper right corner. The beams are shown as the empty circles in the lower left corner in each panel. \label{fig:chan3169}}
\end{figure*}

\begin{figure*}
    \centering
    \includegraphics[width=0.49\textwidth]{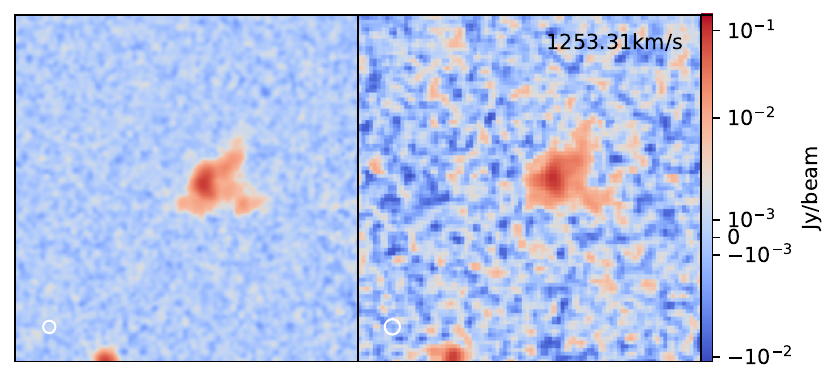}
    \includegraphics[width=0.49\textwidth]{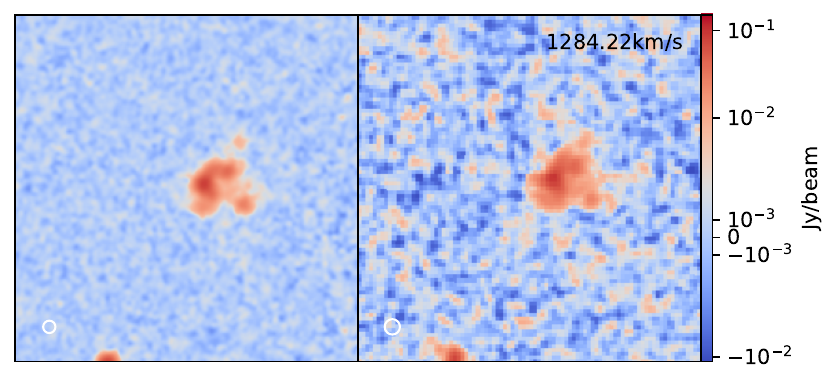}
    \includegraphics[width=0.49\textwidth]{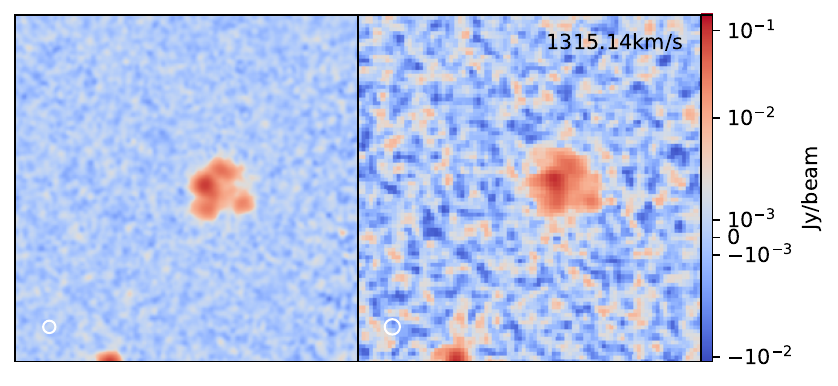}
    \includegraphics[width=0.49\textwidth]{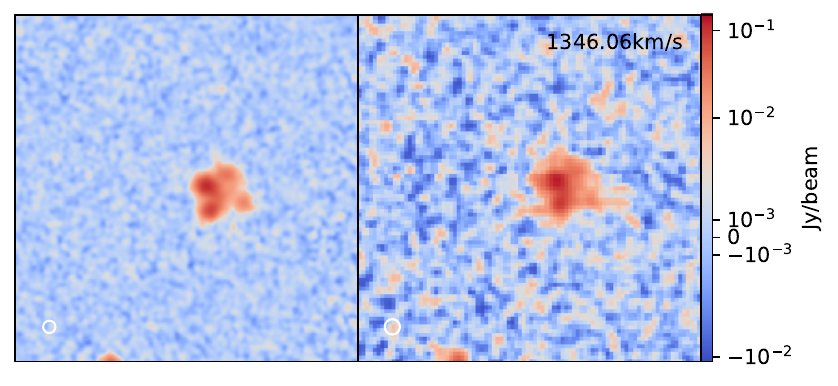}
    \includegraphics[width=0.49\textwidth]{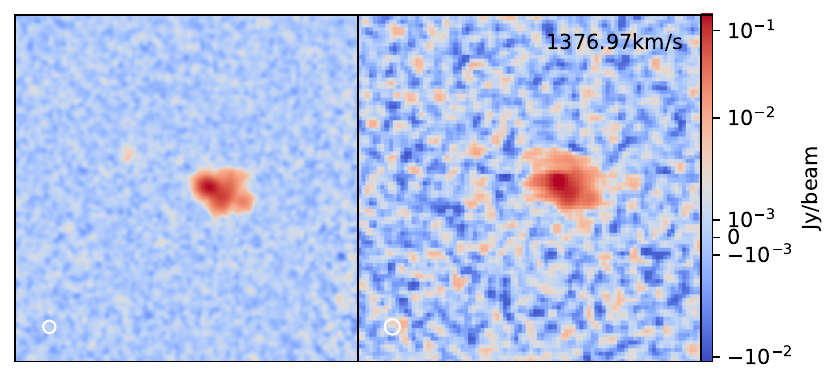}
    \includegraphics[width=0.49\textwidth]{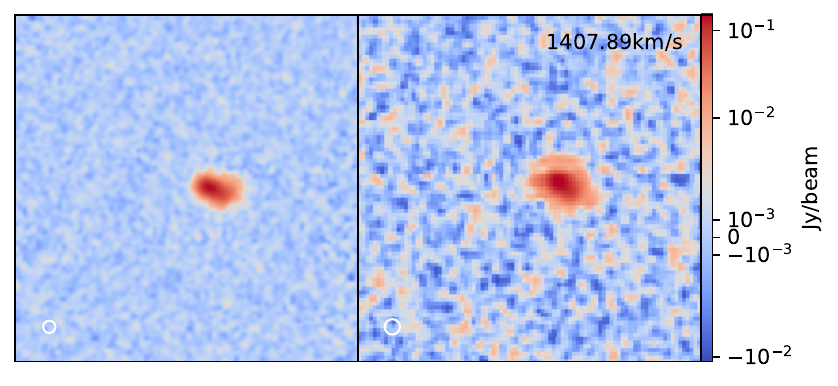}
    \includegraphics[width=0.49\textwidth]{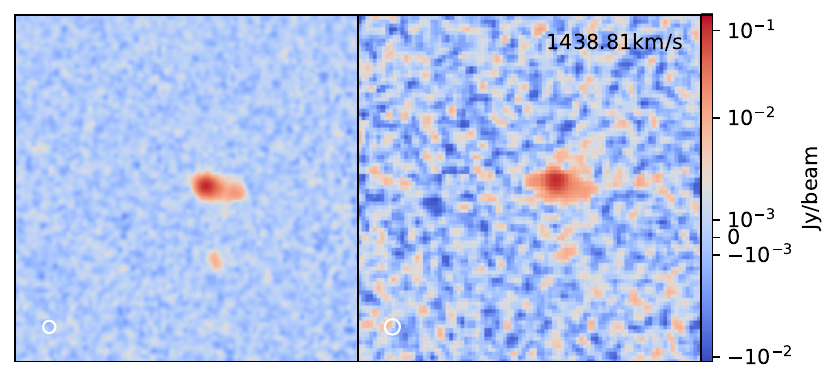}
    \includegraphics[width=0.49\textwidth]{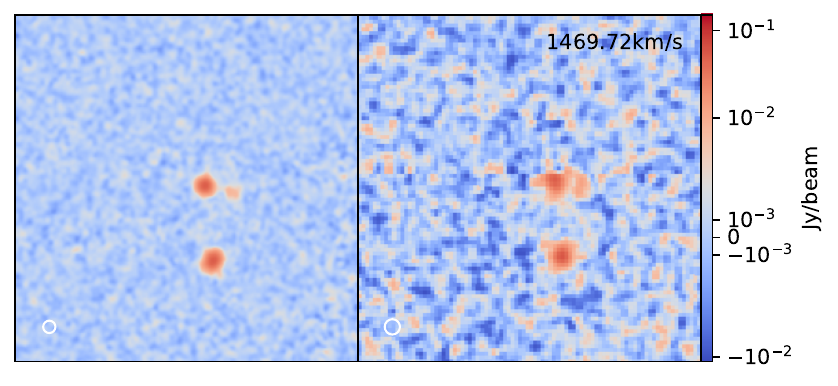}
    \includegraphics[width=0.49\textwidth]{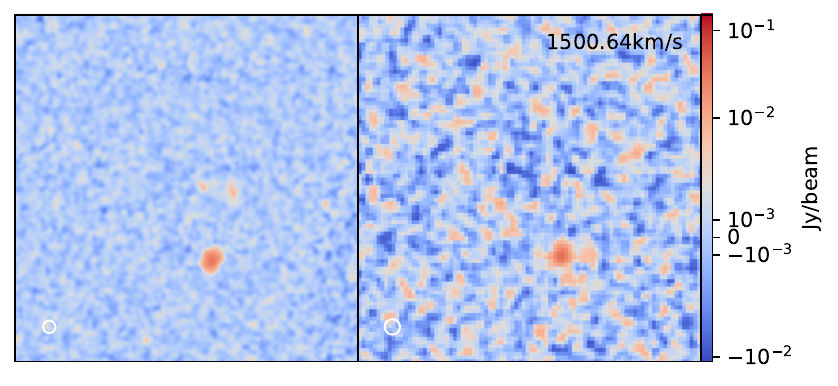}
    \includegraphics[width=0.49\textwidth]{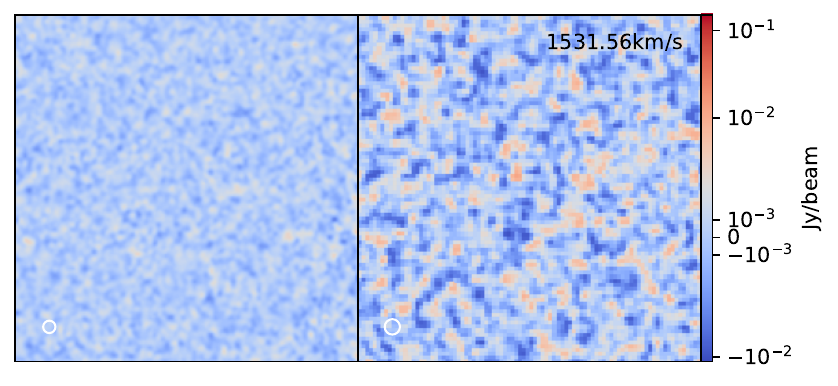}
\contcaption{}
\end{figure*}

\begin{figure*}
    \centering
    \includegraphics[width=0.49\textwidth]{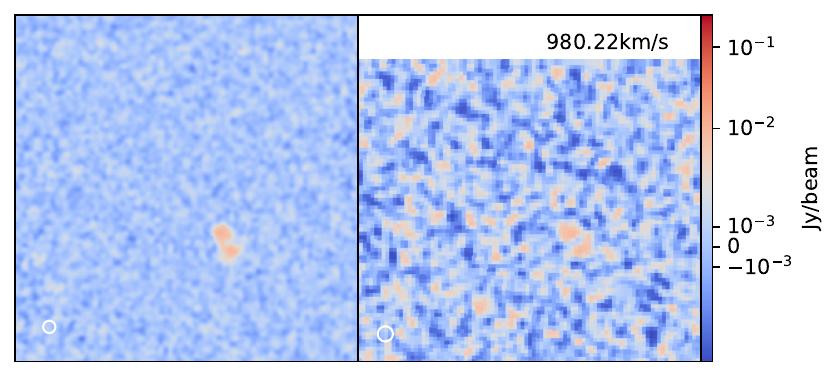}
    \includegraphics[width=0.49\textwidth]{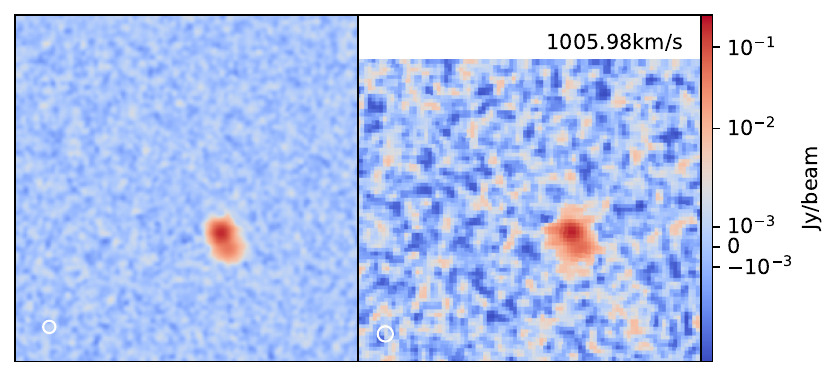}
    \includegraphics[width=0.49\textwidth]{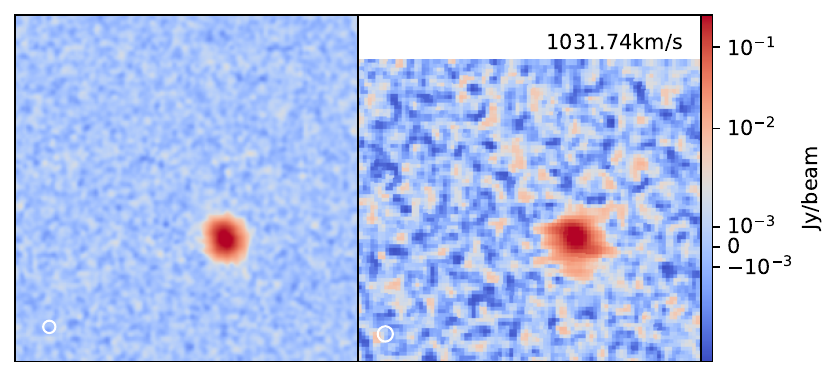}
    \includegraphics[width=0.49\textwidth]{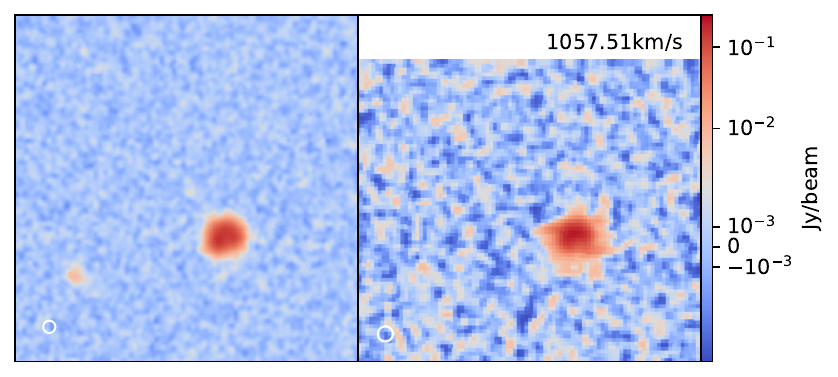}
    \includegraphics[width=0.49\textwidth]{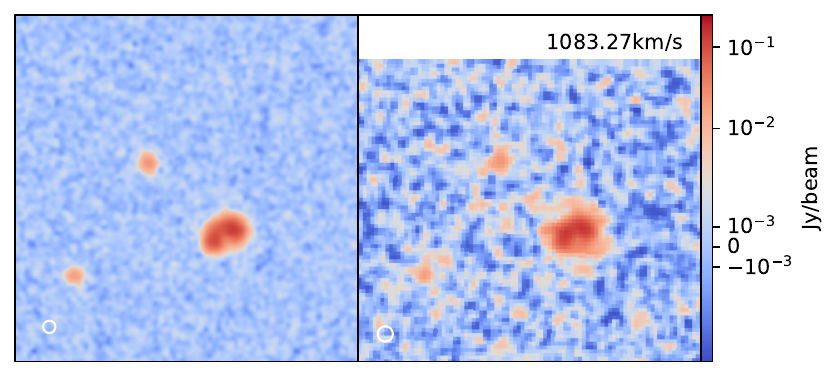}
    \includegraphics[width=0.49\textwidth]{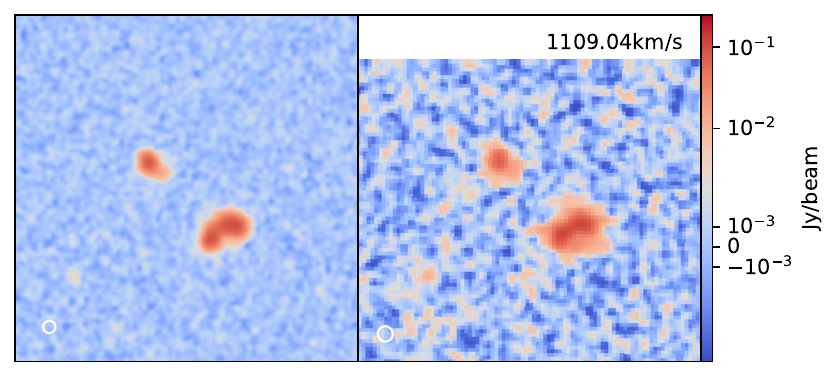}
    \includegraphics[width=0.49\textwidth]{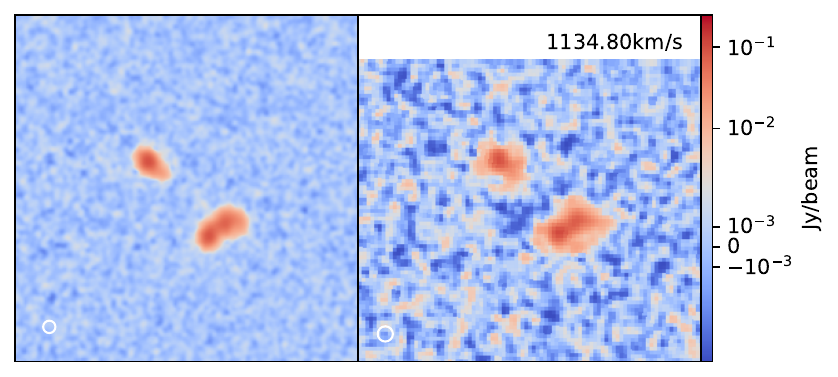}
    \includegraphics[width=0.49\textwidth]{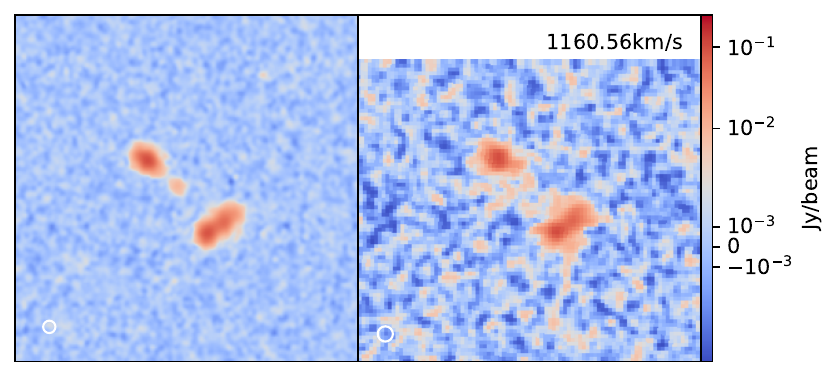}
    \includegraphics[width=0.49\textwidth]{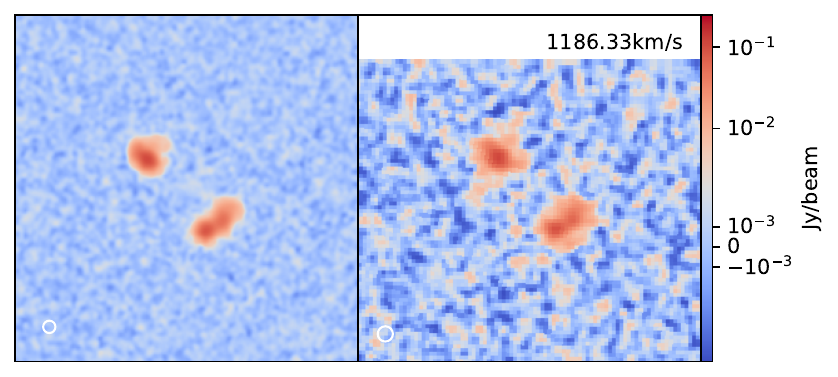}
    \includegraphics[width=0.49\textwidth]{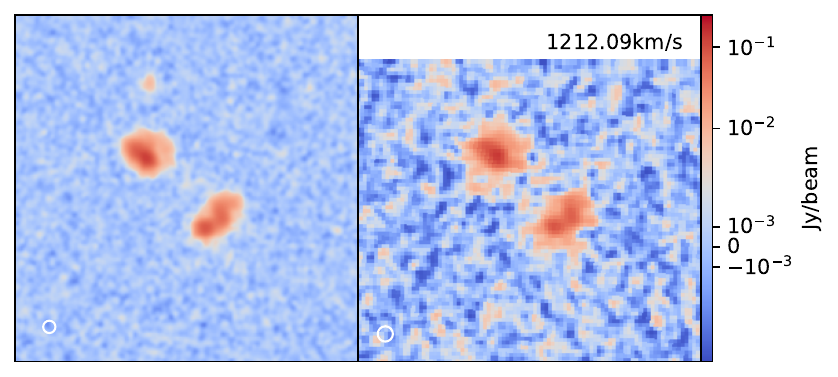}
\caption{The channel maps of the NGC4725-NGC4747 system from FEASTS (left panels) and ALFALFA (right panels). The velocity of the channel is shown in the upper right corner. The beams are shown as the empty circles in the lower left corner in each panel. \label{fig:chan4725}}
\end{figure*}

\begin{figure*}
    \centering
    \includegraphics[width=0.49\textwidth]{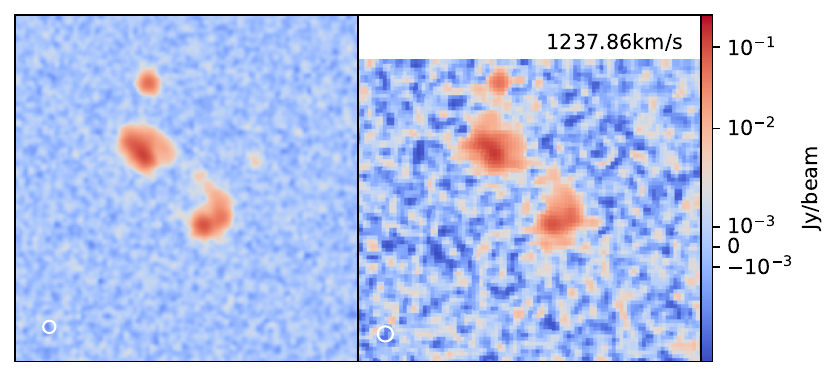}
    \includegraphics[width=0.49\textwidth]{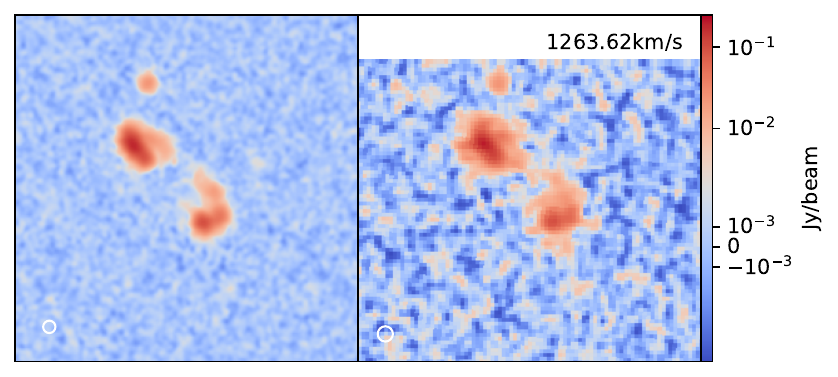}
    \includegraphics[width=0.49\textwidth]{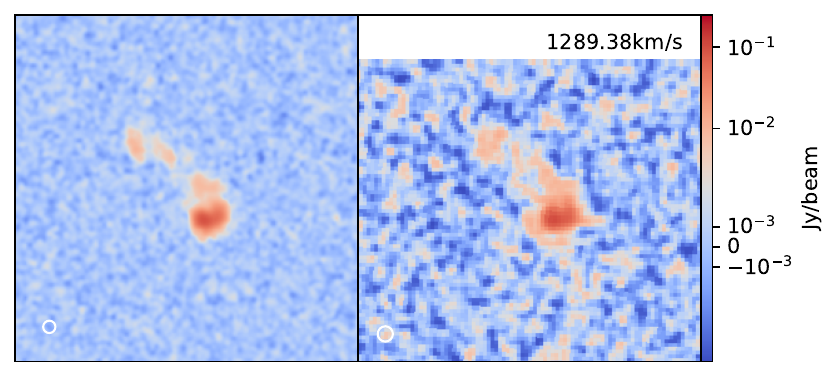}
    \includegraphics[width=0.49\textwidth]{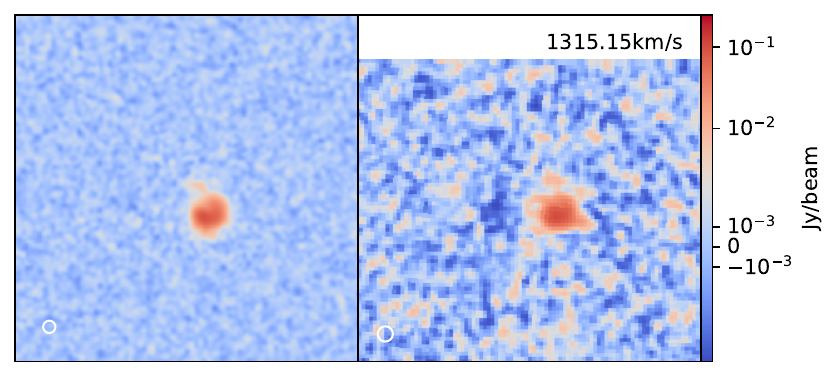}
    \includegraphics[width=0.49\textwidth]{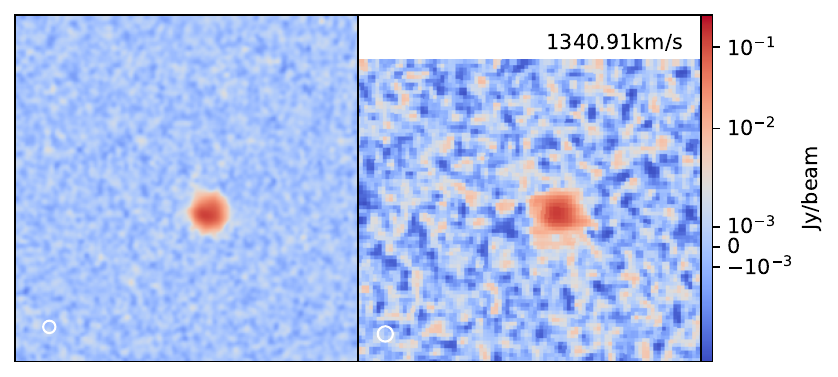}
    \includegraphics[width=0.49\textwidth]{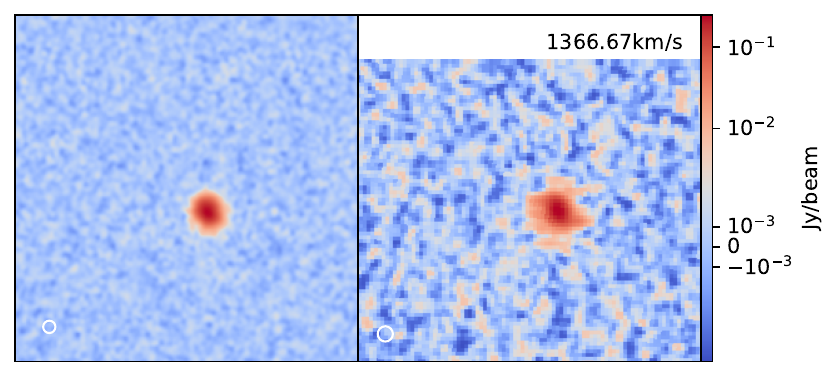}
    \includegraphics[width=0.49\textwidth]{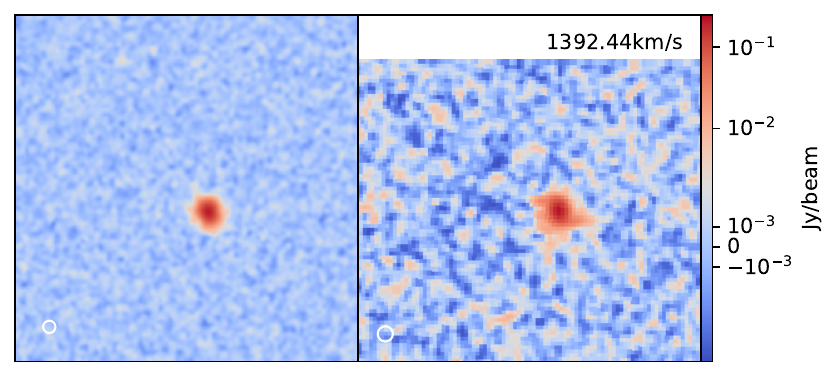}
    \includegraphics[width=0.49\textwidth]{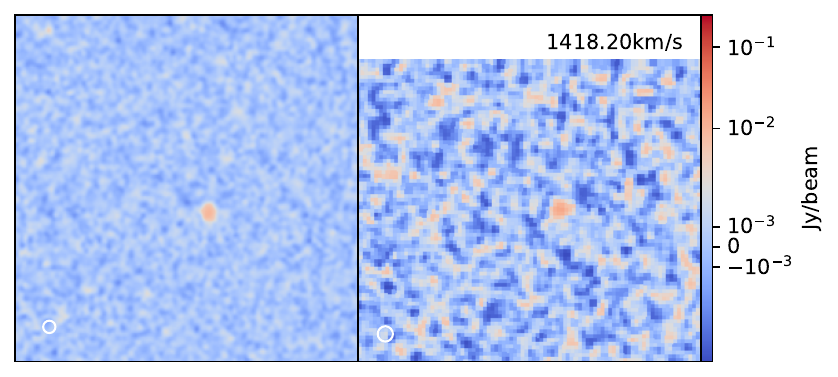}
\contcaption{}
\end{figure*}

\begin{figure*}
    \centering
    \includegraphics[width=0.45\textwidth]{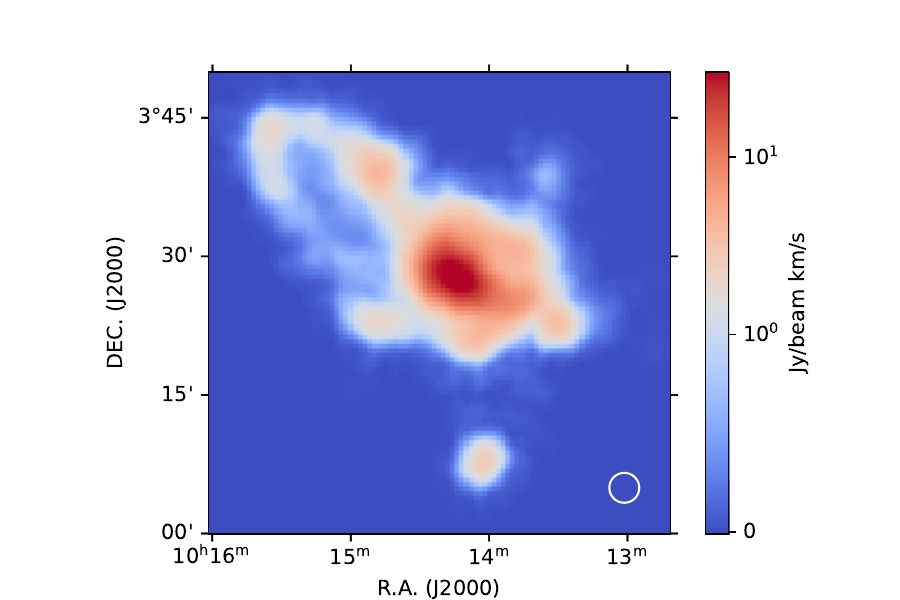}
    \includegraphics[width=0.45\textwidth]{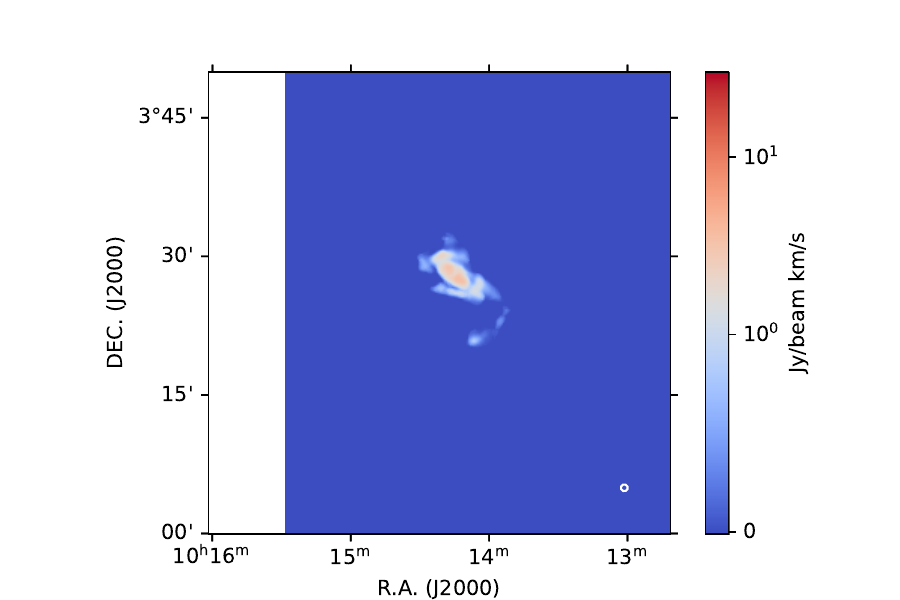}
\caption{The moment-0 maps of the NGC3169-NGC3166 system from FEASTS (left panel) and from GMRT (right panel). The beam is shown as the empty circle in the lower right corner. \label{fig:mom3169}}
\end{figure*}

\begin{figure*}
    \centering
    \includegraphics[width=0.45\textwidth]{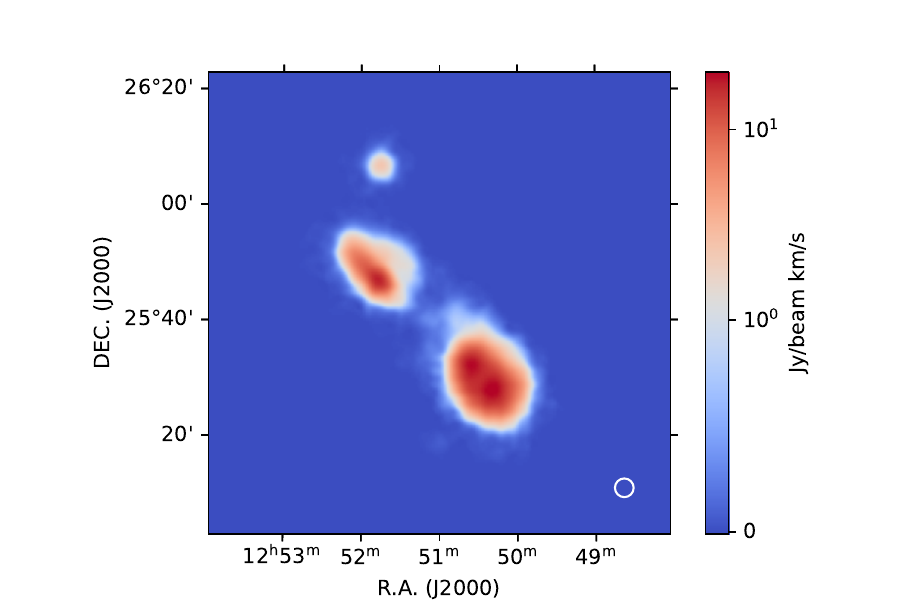}
    \includegraphics[width=0.45\textwidth]{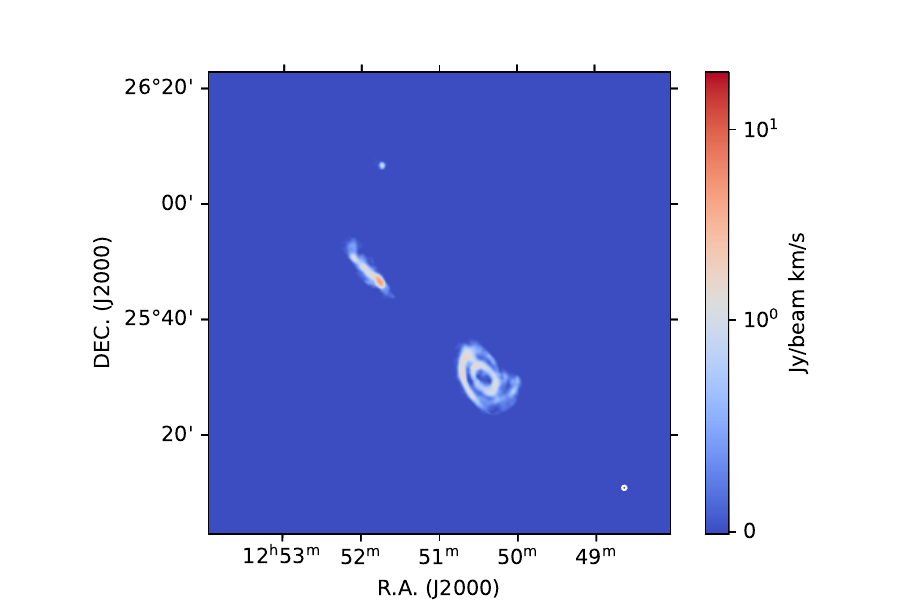}
\caption{The moment-0 maps of the NGC4725-NGC4747 system from FEASTS (left panel) and from GMRT (right panel). The beam is shown as the empty circle in the lower right corner. \label{fig:mom4725}}
\end{figure*}

In Figure~\ref{fig:mom3169} we present the moment-0 map of the NGC3169-NGC3166 system from FEASTS and that from GMRT. The moment-0 maps of the NGC4725-NGC4747 system are shown in Figure~\ref{fig:mom4725}.
The source masks for the GMRT data cubes are constructed using \textit{SoFiA} similarly as for the FEASTS data cubes (Section~\ref{sec:sourceid}). The field-of-view and color scales are identical between different data of the same system.
FEASTS moment-0 maps reveal significantly more \HI{} fluxes and structures at larger galactic radii. They effectively capture the faint and/or large-scale components which are severely missed by GMRT observations. Though lacking small-scale resolution compared to interferometry data, FEASTS moment-0 maps well probe the spatial distribution of \HI{} in tidally interacting galaxies with better completeness.

\section{HI flux division} \label{sec:watershed}
The \HI{} components of small galaxies are identified via visual inspection on both \HI{} moment-0 maps and optical images. We identify one small clump of \HI{} component belong to a small galaxy in each of M77-NGC1055, NGC4725-NGC4747, NGC3169-NGC3166 and NGC5775-NGC5774 systems.

Following \citet{2023ApJ...944..102W} and Huang et al. (in prep.), we use the python function \textit{watershed} from the \textit{skimage.segmentation} module to divide pixels that belongs to the small component from those belong to the major members of the interacting systems. Firstly, we use python function \textit{peak\_local\_max} from \textit{skimage.feature} module to locate local peaks of the \HI{} fluxes in moment-0 maps. Secondly, we adjust the number of peaks to be found so that the small component is located. Then we perform segmentation based on the local peaks via \textit{watershed}. Finally, all segments other than that of the small component are combined. So that the moment-0 maps are segmented into two parts -- main component of \HI{} that are connected to the major members of the interacting system, and the minor component of \HI{} connected to the small galaxies nearby.

We do not perform segmentation in cube domain because the well resolved complex and irregular \HI{} structure in these interacting systems makes it difficult to separate the fluxes.

\section{mock HI profile adjustment} \label{sec:gauss}
The use of \HI{} size-mass relation and the universal \HI{} profile do not automatically produce self-consistent \HI{} disks. I.e., the integrated \HI{} flux of the produced \HI{} profile would not be equal to the \HI{} mass that we input at the beginning.
There are two reasons causing the inconsistency.
The first reason is that the universal profile is derived as the median of the profiles of individual galaxies. This derivation does not guarantee flux conservation.
The second reason is that the slope of the \HI{} size-mass relation is not accurately $0.5$. The non-zero deviation of the actual slope from $0.5$ determines that the \HI{} disks are not strictly self-similar.
Considering that the inconsistency caused by the two facts are both small compared to the large intrinsic uncertainty in the inner region of the universal \HI{} profiles, we decide to adjust the profile in the inner region so that consistency is achieved. The \HI{} size-mass relation, on the other hand, is not changed.

We perform the adjustment as a function of the input \HI{} mass. Firstly, an original \HI{} profile is produced as illustrated in Section~\ref{sec:SDprof}. By integrating the \HI{} profile as a function of radius, we calculate the integrated \HI{} mass of the profile and the deviation from the initial \HI{} mass we input. Then we construct a half-Gaussian centered on $R = 0$ extending only to $R > 0$. The dispersion of it is set to be $\sigma = 0.2 R_{\rm HI}$. And its amplitude is determined so that it integrates to the \HI{} mass deviation. Finally this half-Gaussian is added to the original \HI{} profile and we obtain the adjusted \HI{} profile which integrates to the exactly same amount of \HI{} mass as the input.

The differences of the integrated \HI{} mass between the original and the adjusted \HI{} profile vary from $\sim 2$‰ to $\sim 6 \%$ as a function of the input \HI{} mass.

\section{comparison of HI PSD between interacting systems and mock} \label{sec:PSD}
We compare the \HI{} flux distribution of real and mock galaxies in an order of increasing significance of tidal structures seen in \HI{} PSD (Figure~\ref{fig:PSD}):
\begin{itemize}
    \item We see extraplanar structures beyond the outermost (red, $99.7 \%$) contour of M77-NGC1055, NGC4725-NGC4747 and NGC672-IC1727. Within the outermost contour, the rotation configuration of the main \HI{} disk is largely maintained.
    \item The \HI{} disks of primary and secondary galaxies are clearly connected by a considerable amount of \HI{} in the NGC4631-NGC4656 and NGC660-IC0148 systems. The rotation configuration of the main \HI{} disks is also affected.
    \item The \HI{} in the NGC3169-NGC3166 and NGC5194-NGC5195 systems forms a large envelope surrounding both galaxies. The \HI{} component of the secondary galaxies can not be clearly distinguished from that of the primary galaxy or the envelope. There is also a significant amount of \HI{} outside the envelopes with l-o-s velocities which are mostly consistent with the \HI{} disks.
    \item The observation of the NGC5775-NGC5774 system is limited by the resolution. There is little \HI{} beyond the outermost contour. Projection effect cause the \HI{} components of the two galaxies to overlap with each other significantly. We suspect it is in a similar state to the NGC672-IC1727 system but with weaker tidal interaction.
\end{itemize}

\section{residual maps of the controls} \label{sec:resctrl}
\begin{figure*}
    \centering
    \includegraphics[width=0.39\textwidth]{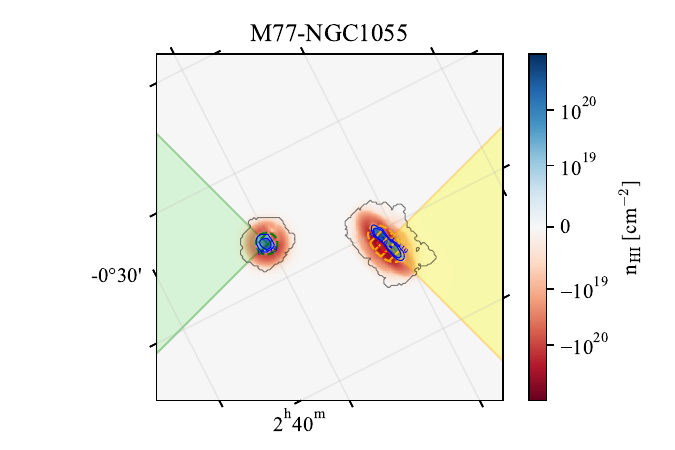}
    \includegraphics[width=0.39\textwidth]{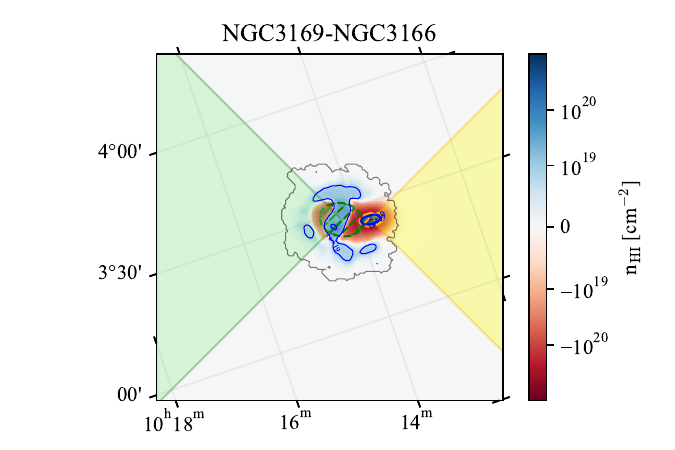}
    \includegraphics[width=0.39\textwidth]{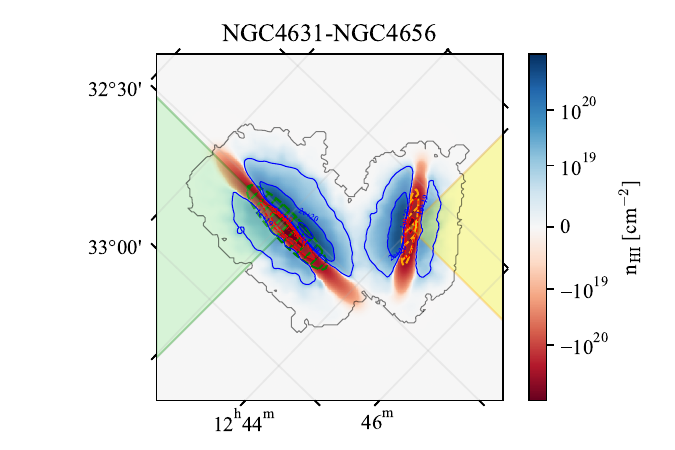}
    \includegraphics[width=0.39\textwidth]{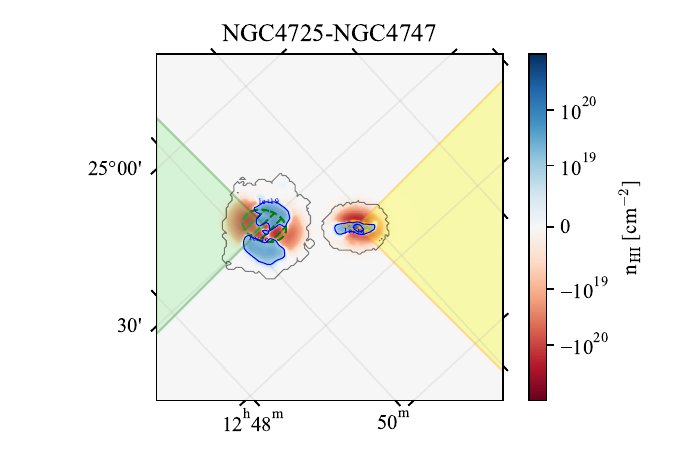}
    \includegraphics[width=0.39\textwidth]{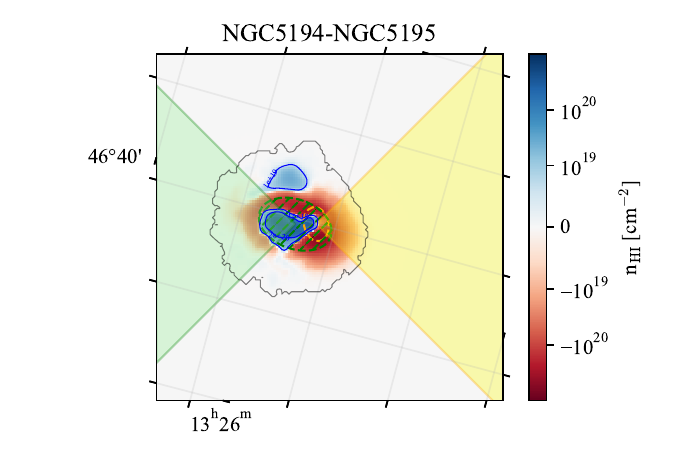}
    \includegraphics[width=0.39\textwidth]{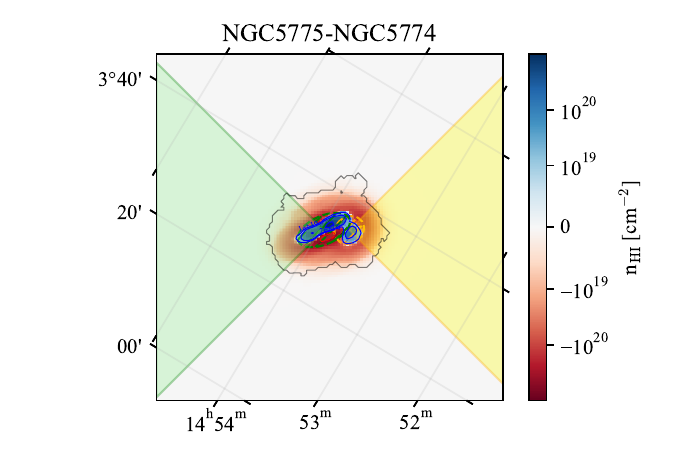}
    \includegraphics[width=0.39\textwidth]{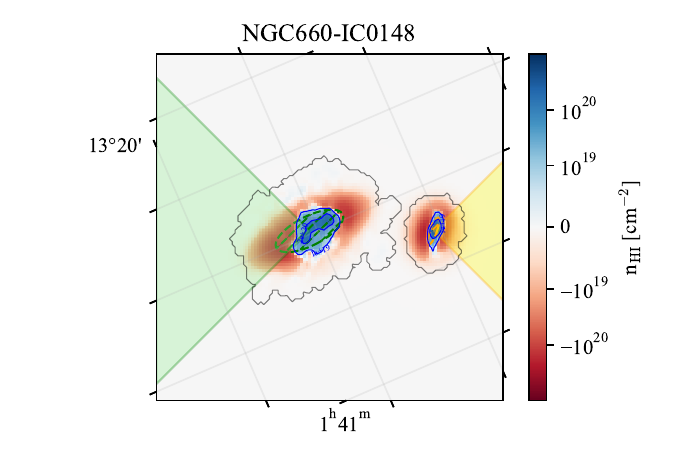}
    \includegraphics[width=0.39\textwidth]{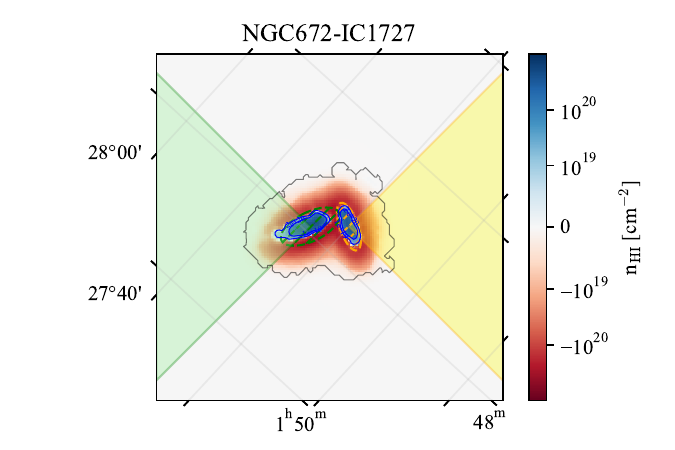}
\caption{The rotated residual maps of the controls. All symbols are the same as those in Figure~\ref{fig:res}. \label{fig:resctrl}}
\end{figure*}

In Figure~\ref{fig:resctrl} we present the residual maps of the controls. For each interacting system, a representative example is shown.

\section{correlations evaluated by Spearman and Kendall tests} \label{sec:spke}
In Table~\ref{tab:corr_sp} we present the Spearman R coefficients and p-values for all the combinations of correlations as in Table~\ref{tab:corr}. And the Kendall $\tau$ coefficients and p-values are presented in Table~\ref{tab:corr_ke}.
For primary galaxies, there is no significant correlation for the \HI{} excess ($\Delta \log M_{\rm HI}$) or corrected \HI{} excess ($\Delta \log M_{\rm HI, cor}$). The only significant correlation is between star formation rate enhancement and $\Delta (\frac{F_{\rm 12}}{F_{\rm 123}})$ based on Spearman test.
For secondary galaxies, both \HI{} excess and corrected \HI{} excess significantly anti-correlate with $\Delta(\frac{R_{\rm 12}}{F_{\rm 123}})$, $\Delta(\frac{F_{\rm 123}}{F_{\rm tot}})$, $\Delta(\frac{R_{\rm 12}}{F_{\rm 12}})$, $\Delta(f_{\rm 12}^+)$, $\Delta(f_{\rm 123}^+)$ and $\Delta (L)$. The star formation rate enhancement also show significant anti-correlations with $\Delta(\frac{R_{\rm 12}}{F_{\rm 123}})$, $\Delta(\frac{F_{\rm 123}}{F_{\rm tot}})$ and $\Delta(\frac{F_{\rm 12}}{F_{\rm tot}})$ based on Spearman tests, and with $\Delta(\frac{F_{\rm 123}}{F_{\rm tot}})$ and $\Delta (\frac{F_{\rm 12}}{F_{\rm tot}})$ based on Kendall tests.
\begin{table*}
\caption{Significance of the correlations between gas content, star formation and \HI{} disorder. The Spearman R coefficient is shown for each correlation, with the corresponding p-value in the following parenthesis. The values for the significant correlations (p-value $< 0.05$) are highlighted in bold. Correlations for primary galaxies are presented on the left while those for secondary galaxies are on the right. \label{tab:corr_sp}}
\begin{tabular}{c|rrr|rrr}
\hline
{} & \multicolumn{3}{c}{primary galaxies} & \multicolumn{3}{c}{secondary galaxies}\\
{} & {$\Delta \log M_{\rm HI}$} & {$\Delta \log M_{\rm HI, cor}$} & {$\Delta \log SFR$} & {$\Delta \log M_{\rm HI}$} & {$\Delta \log M_{\rm HI, cor}$} & {$\Delta \log SFR$} \\
\hline
    $\Delta (\frac{F_{\rm 12}}{F_{\rm 123}})$ & 0.10(0.82)& -0.17(0.69)& \textbf{-0.71(0.05)}& 0.02(0.96)& 0.02(0.96)& -0.50(0.21)
\\
    $\Delta (\frac{R_{\rm 12}}{F_{\rm 123}})$ & -0.52(0.18)& -0.60(0.12)& 0.10(0.82)& \textbf{-0.90(0.00)}& \textbf{-0.90(0.00)}& \textbf{-0.76(0.03)}
\\
    $\Delta (\frac{F_{\rm 123}}{F_{\rm tot}})$ & -0.45(0.26)& -0.57(0.14)& 0.05(0.91)& \textbf{-0.88(0.00)}& \textbf{-0.88(0.00)}& \textbf{-0.83(0.01)}
\\
    $\Delta (\frac{F_{\rm 12}}{F_{\rm tot}})$ & -0.19(0.65)& -0.43(0.29)& -0.33(0.42)& -0.67(0.07)& -0.67(0.07)& \textbf{-0.93(0.00)}
\\
    $\Delta (\frac{R_{\rm 12}}{F_{\rm 12}})$ & -0.62(0.10)& -0.50(0.21)& 0.31(0.46)& \textbf{-0.86(0.01)}& \textbf{-0.86(0.01)}& -0.45(0.26)
\\
    $\Delta (\frac{R_{\rm 123}}{F_{\rm 123}})$ & -0.67(0.07)& -0.40(0.32)& 0.57(0.14)& -0.69(0.06)& -0.69(0.06)& -0.26(0.53)
\\
    $\Delta (f^+_{\rm 12})$ & -0.69(0.06)& -0.60(0.12)& 0.31(0.46)& \textbf{-0.81(0.01)}& \textbf{-0.81(0.01)}& -0.36(0.39)
\\
    $\Delta (f^+_{\rm 123})$ & -0.62(0.10)& -0.48(0.23)& 0.52(0.18)& \textbf{-0.90(0.00)}& \textbf{-0.90(0.00)}& -0.36(0.39)
\\
    $\Delta (S)$ & -0.17(0.69)& -0.33(0.42)& 0.07(0.87)& -0.69(0.06)& -0.69(0.06)& -0.69(0.06)
\\
    $\Delta (L)$ & -0.55(0.16)& -0.48(0.23)& 0.38(0.35)& \textbf{-0.98(0.00)}& \textbf{-0.98(0.00)}& -0.60(0.12)\\
    \hline
\end{tabular}
\end{table*}

\begin{table*}
\caption{Significance of the correlations between gas content, star formation and \HI{} disorder. The Kendall $\tau$ coefficient is shown for each correlation, with the corresponding p-value in the following parenthesis. The values for the significant correlations (p-value $< 0.05$) are highlighted in bold. Correlations for primary galaxies are presented on the left while those for secondary galaxies are on the right. \label{tab:corr_ke}}
\begin{tabular}{c|rrr|rrr}
\hline
{} & \multicolumn{3}{c}{primary galaxies} & \multicolumn{3}{c}{secondary galaxies}\\
{} & {$\Delta \log M_{\rm HI}$} & {$\Delta \log M_{\rm HI, cor}$} & {$\Delta \log SFR$} & {$\Delta \log M_{\rm HI}$} & {$\Delta \log M_{\rm HI, cor}$} & {$\Delta \log SFR$} \\
\hline
    $\Delta (\frac{F_{\rm 12}}{F_{\rm 123}})$ & 0.07(0.90)& -0.07(0.90)& -0.57(0.06)& 0.00(1.00)& 0.00(1.00)& -0.36(0.28)
\\
    $\Delta (\frac{R_{\rm 12}}{F_{\rm 123}})$ & -0.43(0.18)& -0.57(0.06)& 0.07(0.90)& \textbf{-0.79(0.01)}& \textbf{-0.79(0.01)}& -0.57(0.06)
\\
    $\Delta (\frac{F_{\rm 123}}{F_{\rm tot}})$ & -0.36(0.28)& -0.50(0.11)& 0.00(1.00)& \textbf{-0.71(0.01)}& \textbf{-0.71(0.01)}& \textbf{-0.64(0.03)}
\\
    $\Delta (\frac{F_{\rm 12}}{F_{\rm tot}})$ & -0.14(0.72)& -0.43(0.18)& -0.21(0.55)& -0.50(0.11)& -0.50(0.11)& \textbf{-0.86(0.00)}
\\
    $\Delta (\frac{R_{\rm 12}}{F_{\rm 12}})$ & -0.57(0.06)& -0.43(0.18)& 0.21(0.55)& \textbf{-0.64(0.03)}& \textbf{-0.64(0.03)}& -0.43(0.18)
\\
    $\Delta (\frac{R_{\rm 123}}{F_{\rm 123}})$ & -0.50(0.11)& -0.36(0.28)& 0.43(0.18)& -0.57(0.06)& -0.57(0.06)& -0.21(0.55)
\\
    $\Delta (f^+_{\rm 12})$ & -0.57(0.06)& -0.43(0.18)& 0.21(0.55)& \textbf{-0.64(0.03)}& \textbf{-0.64(0.03)}& -0.29(0.40)
\\
    $\Delta (f^+_{\rm 123})$ & -0.43(0.18)& -0.43(0.18)& 0.36(0.28)& \textbf{-0.79(0.01)}& \textbf{-0.79(0.01)}& -0.29(0.40)
\\
    $\Delta (S)$ & -0.14(0.72)& -0.29(0.40)& 0.07(0.90)& -0.50(0.11)& -0.50(0.11)& -0.43(0.18)
\\
    $\Delta (L)$ & -0.43(0.18)& -0.43(0.18)& 0.21(0.55)& \textbf{-0.93(0.00)}& \textbf{-0.93(0.00)}& -0.43(0.18)\\
    \hline
\end{tabular}
\end{table*}


\bsp	
\label{lastpage}
\end{document}